\documentclass[twocolumn,tighten,times]{aastex62}

\newcommand{\pd}{\partial}
\newcommand{\ud}{\ensuremath{\mathrm{d}}}

\newcommand{\kms}{\text{km}}
\newcommand{\kmps}{\text{km}\, \text{s}^{-1}}
\newcommand{\ergsg}{\text{erg}\, \text{g}^{-1}}
\newcommand{\msolar}{\text{M}_{\odot}}
\newcommand{\seconds}{\text{s}}
\newcommand{\kbpnuc}{k_{\text{b}}/\text{nucleon}}
\newcommand{\lmax}{\ell_{\text{max}}}

\newcommand{\kepler}{\textsc{Kepler }}

\usepackage{amsmath,mathtools}
\usepackage{multirow}
\usepackage{ulem}

\received{00, 00, 2019}
\revised{00, 00, 2019}
\accepted{00, 00, 2019}
\submitjournal{ApJ}
\shorttitle{Shell merger in 3D} 
\shortauthors{Yadav et al.}

\begin{document}
\title{Large-Scale Mixing in a Violent Oxygen-Neon Shell Merger Prior to a Core-Collapse Supernova}
\correspondingauthor{Naveen Yadav}
\email{ny@MPA-Garching.MPG.DE}

\author[0000-0002-4107-9443]{Naveen Yadav}
\affil{Max Planck Institute for Astrophysics, 
       Karl-Schwarzschild-Str.~1, D\nobreakdash{-}85748 Garching, Germany}
\affil{Exzellenzcluster ORIGINS, Boltzmannstr.~2, D\nobreakdash{-}85748 Garching, Germany}

\author{Bernhard~M\"uller}
\affil{Monash Centre for Astrophysics, 
       School of Physics and Astronomy, 
       Monash University, Victoria\nobreakdash{-}3800, Australia}
\affil{Astrophysics Research Centre, 
       School of Mathematics and Physics, 
       Queen's University Belfast, Belfast, BT7~1NN, UK}

\author[0000-0002-0831-3330]{Hans Thomas Janka}
\affil{Max Planck Institute for Astrophysics, 
       Karl-Schwarzschild-Str.~1, D\nobreakdash{-}85748 Garching, Germany}

\author{Tobias Melson}
\affil{Max Planck Institute for Astrophysics, 
       Karl-Schwarzschild-Str.~1, D\nobreakdash{-}85748 Garching, Germany}
       
\author[0000-0002-3684-1325]{Alexander Heger}
\affil{Monash Centre for Astrophysics, 
       School of Physics and Astronomy, 
       Monash University, Victoria\nobreakdash{-}3800, Australia}
\affil{Center for Nuclear Astrophysics, Department of Physics and Astronomy, Shanghai Jiao-Tong University, Shanghai\nobreakdash{-}200240, P. R. China}
\affil{Joint Institute for Nuclear Astrophysics, 1 Cyclotron
       Laboratory, National Superconducting Cyclotron Laboratory, Michigan
       State University, East Lansing, MI\nobreakdash{-}48824-1321, USA}

\begin{abstract}
We present a seven-minute long $4\pi$-3D simulation of a shell merger event in a non-rotating $18.88\, \msolar$ supernova progenitor before the onset of gravitational collapse. The key motivation is to capture the large-scale mixing and asymmetries in the wake of the shell merger before collapse using a self-consistent approach. The $4\pi$ geometry is crucial as it allows us to follow the growth and evolution of convective modes on the largest possible scales. We find significant differences between the kinematic, thermodynamic and chemical evolution of the 3D and the 1D model. The 3D model shows vigorous convection leading to more efficient mixing of nuclear species. In the 3D case the entire oxygen shell attains convective Mach numbers of $\mathord{\approx}\, 0.1$, whereas in the 1D model, the convective velocities are much lower and there is negligible overshooting across convective boundaries. In the 3D case, the convective eddies entrain nuclear species from the neon (and carbon) layers into the deeper part of the oxygen burning shell, where they burn and power a violent convection phase with outflows. This is a prototypical model of a convective-reactive system. Due to the strong convection and the resulting efficient mixing, the interface between the neon layer and the silicon-enriched oxygen layer disappears during the evolution, and silicon is mixed far out into merged oxygen/neon shell. Neon entrained inwards by convective downdrafts burns, resulting in lower neon mass in the 3D model compared to the 1D model at time of collapse. In addition, the 3D model develops remarkable large-scale, large-amplitude asymmetries, which may have important implications for the impending gravitational collapse and the subsequent explosion.
\end{abstract}
\section{Introduction}
Multi-dimensional effects in the late burning stages of massive stars have recently garnered considerable interests for a number of reasons. Whereas diffusive mixing models based on the mixing-length theory (MLT) of convection \citep{biermann_32,boehm_58} as used in one-dimensional (1D) stellar evolution models provide a reasonable estimate of mixing \citep[for a more quantitative discussion]{Arnett_2019} in the interior of convective zones for the late, neutrino-cooled burning stages \citep[see Figure~10 \& 14]{mueller_16c}\footnote{The upper panel of Figure~10 shows a comparison between the radial rms velocity 
fluctuations $\delta v_{\rm r}$ (black) in the 3D model to the 
convective velocity $v_{\rm conv}$ computed in the Kepler model 
using MLT (red). Inside the convection zone 
($1.75\,\mathord{-}\,2.25\, \rm M_{\odot}$) the 1D and 3D 
velocities are in close agreement. Figure~14 shows a comparison between the mass fraction profiles in the 3D model and the 1D model. Although, the differences in the mass fractions in the interior of the oxygen shell are minute, the differences in gradients at shell boundaries are conspicuous.}, it has long been speculated that additional phenomena such as turbulent entrainment \citep{fernando_91,strang_01,meakin_07_a,spruit_15} and the excitation of internal waves could significantly affect shell growth, mixing, and angular momentum transport \citep{cantiello_14,fuller_15} in a manner that is not captured by current 1D stellar evolution models.
Convective-reactive systems \citep{Dimotakis_2005} are at the extreme end of mixing problems, where the feedback produced by nuclear burning strongly affects the flow and vice-versa, thereby making it
extremely hard to describe chemical mixing in these systems using simplified 1D prescriptions.

Since the early 1990s, various groups have attempted to study the late convective burning stages using multi-dimensional simulations to investigate such effects. Since the seminal early work in two dimensions (2D; \citealp{bazan_94,bazan_98,asida_00}) and three dimensions (3D; \citealp{kuhlen_03,meakin_06,meakin_07_a,meakin_07_b}), additional convective boundary mixing has indeed been consistently observed in many modern simulations  \citep{mueller_16c,cristini_17,jones_17,andrassy_18} and appears to be well captured by the semi-empirical entrainment laws familiar from terrestrial settings. The long-term impact of such extra mixing on the evolution of massive stars remains more unclear, however. One major caveat concerns the duration of the simulations, which are currently limited to periods far shorter than the thermal adjustment time scale; and it has also been argued that the predicted extrapolation of entrainment rates might in many cases not qualitatively alter stellar structure over secular time scales \citep{mueller_16b}.

In some highly dynamical situations, however, multi-dimensional effects may result in qualitatively different behaviour compared to 1D stellar evolution models.  Such situations often occur when material entrained across shell boundaries burns violently, which can lead to strong feedback on the dynamics of the flow; proton ingestion episodes as studied in 3D by several groups \citep{herwig_11,stancliffe_11,herwig_14,woodward_15} are one noteworthy example. Such violent episodes of convective boundary mixing are also interesting because there is actually a potential fingerprint for the multi-dimensional dynamics in the stellar interior in the form of the nucleosynthesis enabled by turbulent entrainment \citep{herwig_11,jones_16b,ritter_18}.

Such dynamical mixing events may also occur during the late stages of burning in massive stars. 1D stellar evolution models already show that mergers between O, Ne, and C shells are quite commonplace \citep[see, e.g., the $15\, \msolar$ model in their Figure~8]{sukhbold_14}; they are facilitated by relatively small buoyancy barriers between these shells. What is particularly intriguing is that these mergers very often occur only a few turnover time-scales before collapse, as pointed out by the systematic study of \citet{collins_18}, who found such late mergers in about $40$ percent of all progenitors between $16 \, \msolar$ and $26 \, \msolar$. This implies that the merged shells may be caught in a highly dynamical state at the time of the supernova. It is therefore conceivable that traces of such a merger remain imprinted in the explosion geometry. If this is the case, such mergers could help explain asymmetric features in supernova remnants such as the broad Si-Mg rich ``jet''--like features in Cassiopeia~A \citep{delaney_10,isensee_10,grefenstette_14,grefenstette_17}. 1D stellar evolution models with parameterized mixing also suggest that such shell mergers could result in very characteristic nucleosynthesis.

The recent surge of interest in the 3D dynamics of the late burning stages has mostly focused on the potentially beneficial role of convective seed perturbations in the supernova mechanism \citep{mueller_15a,mueller_17,couch_15} and on slow, steady-state entrainment \citep{meakin_07_a,meakin_07_b,mueller_16c,jones_17,cristini_17,andrassy_18}, but such dynamical shell mergers have not yet been investigated thoroughly.  \citet{mueller_16b} reported a late breakout of convection across shell boundaries in a thin O burning shell that was reignited shortly before collapse in an $12.5\, \msolar$ progenitor, but with insufficient time before collapse for a genuine merger of the O and Ne shell. More recently, \citet{mocak_18} observed the merging of shells in a 3D simulation of a $23\, \msolar$ star, but their model was restricted to a small wedge of $27.5^\circ\, \mathord{\times}\, 27.5^\circ$ and  not evolved until core-collapse. Moreover, the merger occurred already within the initial numerical transient before a convective steady state  developed.

Here, we present the first 3D simulation over the full $4\pi$ solid angle of a fully developed large-scale merger of a silicon and neon shell up to the onset of core-collapse. Our simulation follows the last seven minutes of convective O and Ne shell burning in an $18.88\, \msolar$ progenitor \citep{mueller_16a} from an early stage with clearly separated convective shells to the point where the buoyancy jump between the O and Ne shells is reduced sufficiently to trigger a violent merger that is still not fully completed at collapse. In this paper, we focus on analyzing the flow dynamics and mixing during the merger and the comparison of the 3D model to the corresponding 1D stellar evolution calculation. Implications of the shell merger for the ensuing supernova, though a major motivation for the current simulation, will be discussed in future work.

The layout of the paper is as follows. The initial model, which has been derived from a 1D simulation is described in Section~\ref{sec:1DModel}. In Section~\ref{sec:NumericalMethods3DModel}, we provide important details of the numerical method and the microphysics used for the 3D simulation. In Section~\ref{sec:1Dvs3D} we describe in detail the 3D model and compare it with the 1D model. In Section~\ref{sec:pre-sup} we briefly discuss the pre-supernova model. In Section~\ref{sec:Summary} we provide our conclusions and discuss the impact of our results on the pre-supernova structure and the core collapse.
\section{1D Progenitor Model}\label{sec:1DModel}
\begin{figure*}[!]
\begin{center}
\includegraphics[scale=1.0]{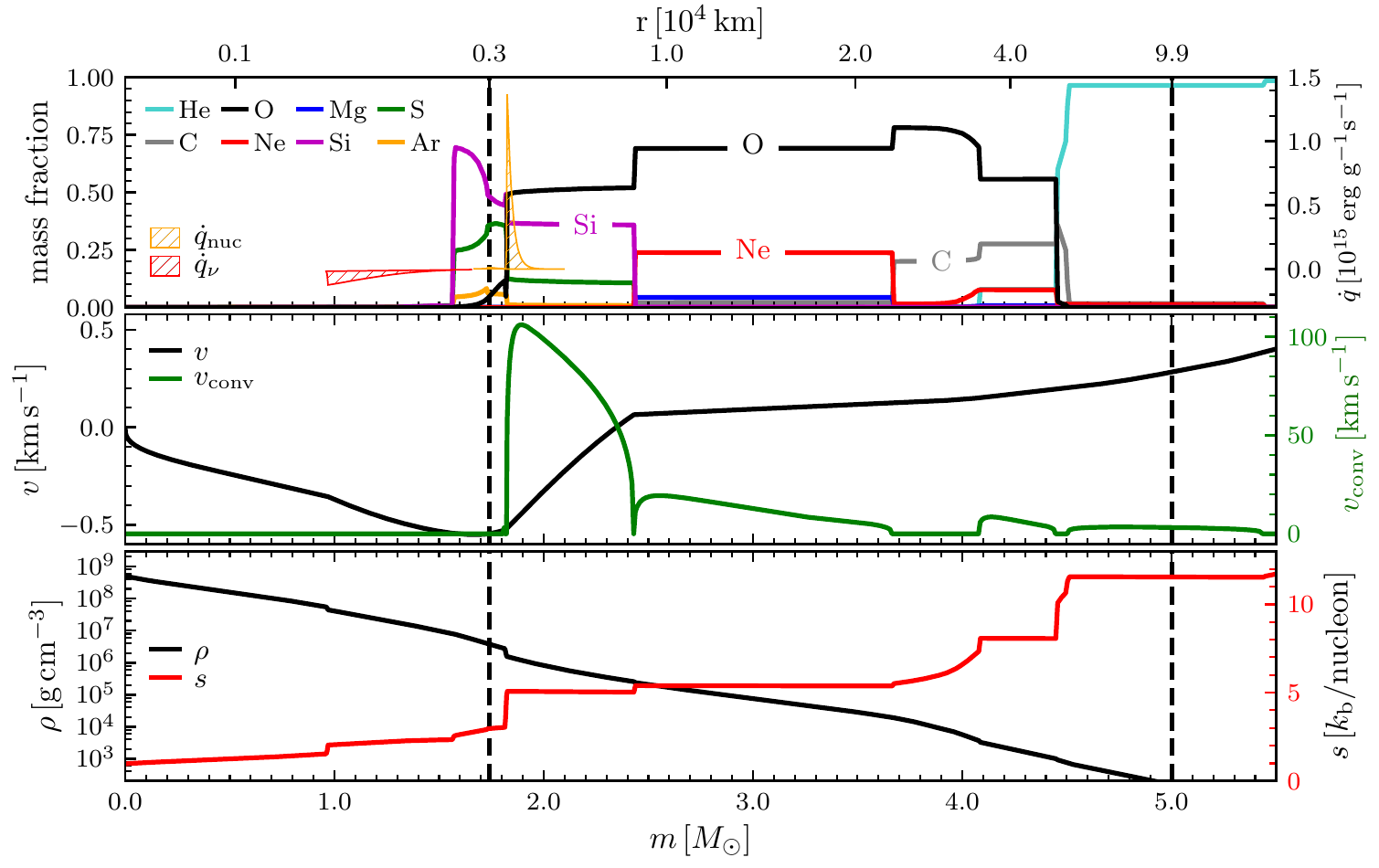}
\end{center}
\vspace*{-9mm}
\caption{Properties of the initial 1D model.
The \textsl{top panel} shows the mass fractions of some relevant $\alpha$-elements, the specific nuclear energy generation rate, and the specific neutrino cooling rate with $|\dot{q}_{\text{nuc}}|\gg |\dot{q}_{\nu}|$ at the base of the O burning shell, whereas $|\dot{q}_{\nu}|\gg |\dot{q}_{\text{nuc}}|$ in the Fe/Si core. The \textsl{middle panel} shows the radial velocity and the convective velocity according to MLT. The Si/Fe core is slowly contracting whereas the O shell is slowly expanding. The O burning shell has larger convective velocity compared to the Ne and C burning shells, and the Fe/Si core is convectively inert. The bottom panel shows the density and entropy profiles. Entropy jumps correspond to composition interfaces; large relative jumps in entropy means high stiffness of such boundaries against convective entrainment. Vertical black dashed lines mark the shell which is simulated in 3D.}\label{fig:figure_01}
\end{figure*}
We consider a progenitor with
solar metallicity and a zero-age main sequence mass of $18.88\, \msolar$. The 1D model has been calculated using the \kepler stellar evolution code \citep{heger_10} as outlined in \citet[their Section~2.1]{mueller_16c}\footnote{
\kepler uses the Ledoux criterion for semiconvection as described in \citet{weaver_78}.
We use $\alpha\,\mathord{=}\,1$ for MLT (larger [$\mathord{\approx}\,1.7$] and depth-dependent MLT values 
used in the solar convection zone may only be valid for the surface convection 
zone of the sun).
Convective boundaries are `softened' by one 
formal overshoot zone with $1/10$ mixing efficiency as 
semiconvection \citep{weaver_78}.
We emphasize that we do not use the $f$-parameter model for overshoot in our 1D models, as
the physics behind ``overshooting prescriptions'' is very uncertain, and there is no
evidence or consensus about their validity in all evolution stages.
}.
The 1D model is mapped to the 3D grid $420\, \seconds$ before the onset of core collapse. Figure~\ref{fig:figure_01} shows the relevant properties of the initial 1D model. The top panel shows the mass fraction profiles for key nuclei, the specific nuclear energy generation rate, and the specific neutrino cooling rate. The shells of principal interest in this paper are the O-rich shells between $1.7\mathord{-}4.5\, \msolar$ ($\mathord{\approx}\, 2,000\mathord{-}42,000\, \kms$). These include a convective O burning shell\footnote{Following common practice in stellar evolution literature, we label the convective shells by the fuel that burns at their base, unless we explicitly discuss the composition in more detail.} with ashes of Si and S, a convective Ne burning shell between $2.5\mathord{-}3.7\, \msolar$, which is separated by a non-convective layer from a convective C burning shell with ashes of O and Ne between $4.1\mathord{-}4.5\, \msolar$. The O and Ne shells are separated by a thin interface at $\mathord{\approx}\, 2.45\, \msolar$. The Si shell and the Fe core inside $1.6\, \msolar$ are practically inert, with neutrino cooling dominating over burning. The middle panel shows the profiles of the radial velocity, $v$, and the convective velocity, $v_{\text{conv}}$, calculated using the same choice of dimensionless coefficients as \citep{mueller_16b}. The convective velocity near the base of the O shell reaches up to $\mathord{\sim}\, 100\, \kmps$. The Ne and C shells are relatively quiet, having  convective velocities close to $20\, \kmps$ and $10\, \kmps$, respectively. The bottom panel displays the density and (specific) entropy profiles. The entropy profile shows the three aforementioned convective regions as flat sections. Significant entropy jumps of $\mathord{\approx}\, 2.0\, \kbpnuc$ and $\mathord{\approx}\, 4.0\, \kbpnuc$ (here $k_{\text{b}}$ is the Boltzmann constant) close to the bottom (just above the Fe/Si core) and close to the top (just below the He shell) of the three shells make the convective boundaries quite stiff, with the inner boundary of the oxygen shell even being visible as a discontinuity in the density profile.  By contrast, the entropy jump between the oxygen and neon shells is small, and the neon and carbon shells are separated by a stable region with gradually increasing entropy, but not by a discontinuity.

To be clear, choosing ``$18.88\, \msolar$'' (accurate to two decimal places) does \textit{not} amount
to ``fine-tuning'' of the 
model presented in \citet{mueller_16c}. We would like to emphasize that the stellar model presented 
in this work and the stellar model used by \citet{mueller_16c} 
are $\mathord{\approx}\, 1\rm \, \msolar$ apart, and it is well known 
that the pre-supernova structure of a star is a very sensitive 
function of the zero-age main sequence mass (ZAMS)
\citep[see upper left panel of their Figure~8]{Sukhbold_2018}. 
 \citet{collins_18} evaluated convective velocities and eddy 
scales in the oxygen and silicon burning shells of a large set 
(2353 solar-metallicity, non-rotating, calculated using the \kepler code) 
of progenitor models with ZAMS between $9.45$ and $35\rm \msolar$. 
One of their key finding is that shell mergers are not rare; 
``about 40 percent of progenitors between 16 and 26 $\rm \msolar$ 
exhibit simultaneous oxygen and neon burning in the same convection 
zone as a result of a shell merger shortly before collapse''. At the same time, there 
is no clear mapping between the outcomes in 1D and 3D, and therefore, 
it cannot be said \emph{a priori} with certainty if a certain 1D model 
will exhibit a shell merger in 3D simulation.
In a nutshell, the $18.88\, \msolar$ model was 
chosen for this study as the dominant angular wave numbers ($\ell$) of convection in Si and O shell 
were small, and at the same time the convective Mach number was large.
\section{Numerical Methods: 3D Model}\label{sec:NumericalMethods3DModel}
\begin{deluxetable}{l l}
  \tablecaption{Setup details for the 3D simulation.\label{tab:data-struct}}
  \tablewidth{0pt}
  \tablehead{Quantity & Option/Value}
\startdata
\tablenotemark{a}Radial grid                                      & Geometric\\
Cells in radial direction, $N_{r}$    & 450\\
\tablenotemark{b}Cells in polar direction, $N_{\theta}$    & 48\\
\tablenotemark{b}Cells in azimuthal direction, $N_{\phi}$    & 148\\
\tablenotemark{c}No. of ghost cells ($\theta,\, \phi$), $N_g$             &    8\\
\tablenotemark{c}No. of buffer cells ($\theta,\, \phi$), $N_b$            &    2\\
Angular resolution                                    & $2^{\circ}$\\
\tablenotemark{d}Inner boundary at $t=0$, $r_{-}$ & $3.30\times 10^3\, \kms$ \\
\tablenotemark{d}Outer boundary at $t=0$, $r_{+}$ & $9.88\times 10^4\, \kms$ \\
Inner boundary condition, radial & refer to Section~\ref{subsec:bcs}\\
Outer boundary condition, radial & refer to Section~\ref{subsec:bcs}\\
Gravitational potential          & Newtonian, Spherical\\
Simulation time                  & $420\, \seconds$\\
\tablenotemark{e}Perturbation amplitude ($f$)     & $2\times 10^{-4}$\\
Neutrino cooling                 & Yes\\
Equation of state                & Helmholtz EoS\\
\tablenotemark{f}CFL             & 0.6\\
\enddata
\tablenotetext{a}{$r_{i+1}=(1+\alpha)r_{i}$, where $1+\alpha = \left(r_{+}/r_{-}\right)^{1/N_r}$.}
\tablenotetext{b}{The number of angular cells are specified per patch of the  Yin-Yang grid.}
\tablenotetext{c}{For each grid patch (Yin and Yang).}
\tablenotetext{d}{The inner and outer boundaries are moved according to the trajectory obtained from 1D model.}
\tablenotetext{e}{$\delta \rho  = f\rho\, U[-1,1]$, where $U[-1,1]$ is a uniformly distributed random number in the interval $[-1,1]$.}
\tablenotetext{f}{CFL is the value of the Courant--Friedrichs--Lewy condition.}
\vspace{-3.5em}
\end{deluxetable}
The 3D simulation uses the \textsc{Prometheus} code. The simulation employs a moving (but non-Lagrangian) grid in the radial direction and an axis-free Yin-Yang grid \citep{kageyama_04,wongwathanarat_10a} as implemented in \citet{melson_15a} with a uniform angular resolution ($2^{\circ}$) in the $\theta$ and $\phi$ directions\footnote{\textit{Adequacy} of $2^{\circ}$ angular resolution: A simulation using an equidistant spherical grid ($\Delta\theta\,\mathord{=}\,\Delta\phi\,\mathord{=}\,\alpha$) captures angular modes with $\ell$ ranging between zero and $180^{\circ}/\alpha$. Therefore, in our case with $\alpha\,\mathord{=}\,2^{\circ}$, we can extract modes up to 
$\ell\,\mathord{=}\,90$. The well developed inertial 
range (see left panel of Figure~15 in Section~\ref{sec:pre-sup}) is a proof 
that the simulation is in the regime of implicit large eddy simulations 
\citep[ILES]{Boris_1992}, and the energy transfer between different 
length scales is correctly modelled. Moreover, ignoring the smallest scale does 
not affect the evolution, as the mixing (and the related hot-spot burning as described in Section~\ref{sec:TE}) is 
caused by the large-scale convective flow.}. For the purpose of this simulation we changed the $4^{\mathrm{th}}$-order reconstruction originally used in the piecewise-parabolic method (PPM) of \citet{colella_84} to a $6^{\mathrm{th}}$-order extremum-preserving reconstruction \citep{colella_08,sekora_09}. We use the Helmholtz equation of state (EoS) described in \citet{timmes_99} and \citet{timmes_00}, which is thermodynamically consistent to high accuracy. Nuclear burning is treated using a 19-species $\alpha$-network
\citep{weaver_78}. An accurate modeling of silicon burning is avoided as its nuclear burning in the quasi-statistical equilibrium (QSE) regime involves tracking a very large number of nuclear species making it computationally expensive. Energy loss by neutrinos \citep{itoh_96} is included as a local sink term. 
\subsection{Boundary Conditions}\label{subsec:bcs}
We excise the core inside $1.7\, \msolar$ as most of the nuclear energy generation is limited to a small region at the base of the O shell (described in Section~\ref{sec:1DModel}) and the Si/Fe core is relatively inert. The region outside of $5.0\, \msolar$ is also excluded from the simulation as it remains dynamically disconnected from the interior during the short 3D simulation. In the 1D model the Si/Fe core cools via neutrino emission and contracts. For consistency, the radial boundaries of the computational domain are moved in step with the 1D model (movement of the outer boundary is negligible in practice). Thus, the 3D simulation covers 1) a small part of the Si shell as a ``buffer'' below the O shell, 2) the entire O, Ne, and C shells as the ``region of interest'' layers, and 3) a small part of the He shell. We emphasize that the steep entropy step at the Si/O interface is about $0.1\, \msolar$ away from the inner grid boundary (see Figure~\ref{fig:figure_01}). This convectively stable interface ensures that the convective mass motions in the simulation volume remain radially separated from the grid boundary.
\phantomsection \label{para:bcs}
For PPM reconstruction, we impose reflective boundary conditions 
for velocity at both the inner and outer radial boundary and extrapolate 
all thermodynamical quantities assuming adiabatic and hydrostatic 
stratification. We strictly enforce zero advective fluxes across the 
boundaries before updating the conserved variables.
\subsection{1D-to-3D Mapping}
The 3D simulation is initialized using density, temperature, and 
mass fraction profiles taken from the 1D model. Hydrodynamic and 
thermodynamic variables are mapped using second order polynomial interpolation. 
Mass fractions are mapped using linear interpolation (to preserve the 
discontinuities at shell boundaries). The question of initial transients (mapping errors) is discussed  in Section~\ref{sec:transients}. 
In order to break the spherical symmetry we introduce small
deviations from hydrostatic equilibrium by perturbing the density 
field such that $\delta \rho  = U[-1,1]f\rho$, where $U[-1,1]$ is a uniformly distributed random number in the interval $[-1,1]$ and the amplitude $f=2.0\times 10^{-4}$.

The Newtonian gravitational potential is computed in the monopole approximation; the appropriate contribution from the excised inner core is added. The details of the simulation setup are summarized in Table~\ref{tab:data-struct}.
We have not done a convergence study for the model presented in the paper. 
Please refer to the discussion on the effect of angular and radial 
resolution on a simulation of oxygen burning in case of the $18\, \rm \msolar$ model by \citet[see Appendix~B]{mueller_16c}. The simulations are converged in terms of the coupling between nuclear energy generation rate and growth of turbulent 
motions \citep[Figure~18]{mueller_16c}; and dominant multipoles \citep[$\ell$-modes in Figure~19]{mueller_16c} emerging in the flow.
\section{Simulation Results: 1D vs 3D}\label{sec:1Dvs3D}
In the following sections we discuss the kinematic, thermodynamic, and 
chemical evolution of the 3D model and compare it to the 1D model.
We start by describing the convective instability of the 
simulated shell as a function of time.
\subsection{Convective stability}\label{subsec:conv-stab}
\begin{figure*}[!]
\vspace*{5mm}
\begin{center}
\includegraphics[width=\linewidth]{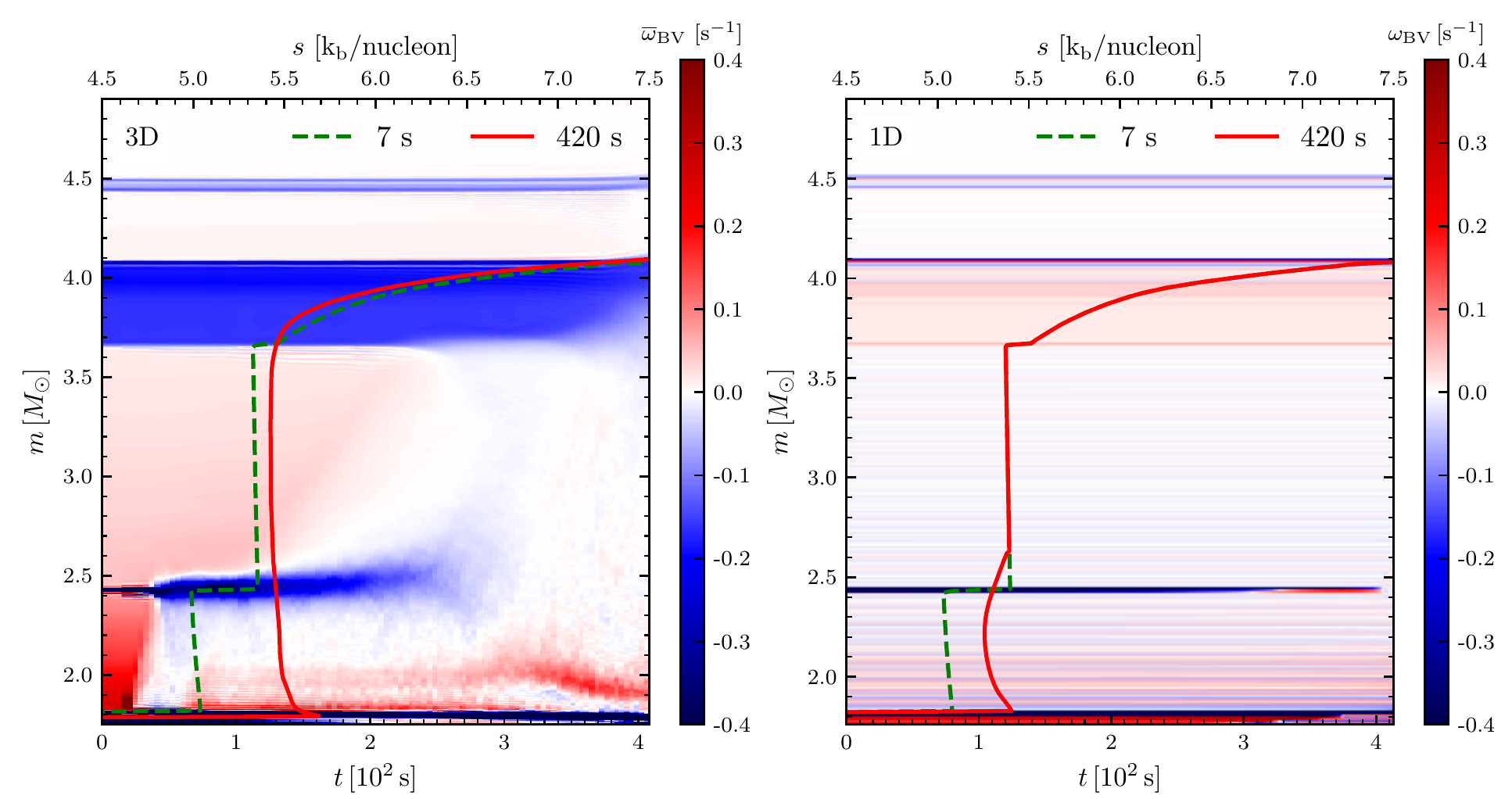}
\end{center}
\vspace*{-6mm}
\caption{Left panel: Space-time plot showing the spherically averaged Brunt-V{\" a}is{\" a}l{\" a} frequency in the 3D run. We use the definition of Brunt-V{\" a}is{\" a}l{\" a} frequency from \cite{buras_06a}. According to this definition, regions with negative $\omega_{\text{BV}}$ are convectively stable and regions with positive $\omega_{\text{BV}}$ are convectively unstable according to the Ledoux criterion.
Initially the active O shell ($1.8\mathord{-}2.4\, \msolar$),
the active Ne shell ($2.4\mathord{-}3.7\, \msolar$),
and the active C shell ($4.1\mathord{-}4.4\, \msolar$)
are clearly separated by stable layers with
positive gradients in the initial entropy profile
(green curve for entropy vs. mass see axis on top of panel). The barrier between the O
and Ne shells at $2.5\, \msolar$  disappears
after  $\mathrm{250}\,\seconds$ due to the increasing
entropy in the O shell. The result is a large
merged convective layer with a slightly negative
entropy gradient between $\mathord{\approx}\, 1.8\mathord{-}3.7\, \msolar$ at collapse (red curve for entropy vs. mass see axis on top of panel).
Right panel: Space-time plot showing the Brunt-V{\" a}is{\" a}l{\" a} frequency for the 1D model. The convectively stable interfaces in the 1D model are very narrow. The stability properties in most of the O shell remain unchanged during the course of evolution, except the changes in convectively stability at the base of  Zone~I close to the end (sign reversal).}
\label{fig:figure_02}
\end{figure*}
The Ledoux criterion can be used to test the stability of a mass element against convective overturn. The Brunt-V{\" a}is{\" a}l{\" a} frequency is given by
\begin{equation}
\omega_{\text{BV}}^2  = \left(\frac{\pd \ln \rho}{\pd r}
-\frac{1}{\Gamma_{1}}\frac{\pd \ln p}{\pd r}\right)g,\label{eqn:bv}
\end{equation}
where $g\,\mathord{=}\,-\ud\Phi/\ud r\,\mathord{<}\,0$ is 
the gravitational acceleration and 
\begin{equation}
\Gamma_1 = \left(\frac{\pd \ln p}{\pd \ln \rho}\right)_{s,X_i},   
\end{equation}
is the adiabatic index. For unstable modes $\omega_{\text{BV}}\,\mathord{>}\,0$ is the growth rate; 
for stable modes $\omega_{\text{BV}}$ is imaginary and $|\omega_{\text{BV}}|$ is the oscillation frequency.
Following \citet{buras_06a}, we redefine the Brunt-V{\" a}is{\" a}l{\" a} frequency such that
\begin{eqnarray}
\omega_{\text{BV}} &\coloneqq& \mathrm{sign}(C_\mathrm{L}) \sqrt{-\frac{g}{\rho}
 |C_\mathrm{L}|},\label{eq:omega_BV_mod}\\
C_\mathrm{L} &\coloneqq&  \left(\frac{\pd \rho}
{\pd s}\right)_{X_i,p} \frac{\ud s}{\ud r}
+\sum_i\left(\frac{\pd \rho}{\pd X_i}\right)_{s,p} \frac{\ud X_i}{\ud r},
\label{eq:omega_BV}
\end{eqnarray}
which is more convenient for visualization purposes.
The conversion of Equation~\eqref{eqn:bv} to Equation~\eqref{eq:omega_BV} is exemplified in Appendix A of \citet{mueller_16c}\footnote{\citet{heger_05} show an alternative approach (used in \kepler stellar evolution code) which can be used to extract the contribution of composition to $\omega_{\text{BV}}$ without explicitly tracking composition.}. Note that our definition changes only affects the sign convention, but not the absolute value of $\omega_{\text{BV}}$.
For unstable modes ($C_\mathrm{L}\, \mathord{>}\,0$), $\omega_{\text{BV}}$ is the growth rate, and for stable modes ($C_\mathrm{L}\, \mathord{<}\, 0$), $\omega_{\text{BV}}$ is the negative of the oscillation frequency. Therefore, according to the sign convention adopted in this paper $\omega_{\text{BV}}\, \mathord{>}\, 0$ corresponds to convective instability.
\begin{deluxetable}{l c c c r}
  \tablecolumns{5}
    \tablecaption{Convectively stable zones according to the Ledoux criterion at the beginning of the 3D simulation. \label{tab:stablezones}}
  \tablehead{\colhead{Zone}   & \multicolumn{2}{c}{Location (inner edge)} & \multicolumn{2}{c}{Width}\tabularnewline
  \colhead{}                  &\colhead{$m\, [\msolar]$} & \colhead{$r\,[\kms]$}  & \colhead{$\Delta m\,[\msolar]$} & \colhead{$\Delta r\,[\kms]$}}
\startdata
  I & 2.4 & $\ 7,500$ & 0.1 & $1,000$ \tabularnewline
  II\tablenotemark{a} & 3.7 & $21,500$ & 0.4  & $10,000$ \tabularnewline
  III & 4.4 & $49,000$ & 0.1  & $5,000$
\enddata
\tablenotetext{a}{Zone~II in the 1D model shows a strong and narrow barrier ($\mathord{<}\, 0.1\, \msolar$) close to $\mathord{\approx}\, 4.1\, \msolar$.}
\vspace{-6mm}
\end{deluxetable}

Figure~\ref{fig:figure_02} (left panel) shows the spherically averaged Brunt-V{\" a}is{\" a}l{\" a} frequency $\overline{\omega}_{\text{BV}}$ profiles as a function of time for the 3D model. $\overline{\omega}_{\text{BV}}$ is defined as
\begin{equation}
    \overline{\omega}_{\text{BV}}(r) \coloneqq \frac{1}{4\pi}\int\limits_{\Omega} \omega_{\text{BV}}(\mathbf{r})\, \ud \Omega,\label{eqn:avg_brunt_vaisala}
\end{equation}
which is different from $\omega_{\text{BV}}$ calculated using spherically averaged $\rho$, $p$ etc. The plot shows three convectively stable layers (zones~I$-$III at masses of $2.4\mathord{-}2.5\, \msolar$, $3.7\mathord{-}4.1\, \msolar$, and $4.4\mathord{-}4.5\, \msolar$ respectively) sandwiched between more voluminous convectively unstable regions. The location ($m$ and $r$) and width ($\Delta m$ and $\Delta r$) of these zones at $t=0$ are listed in Table~\ref{tab:stablezones} in order of increasing mass coordinate. The initial and final entropy profiles (green and red lines in left panel of Figure~\ref{fig:figure_02}) show the disappearance of entropy jumps between the convectively stable and unstable regions. Such entropy steps represent barriers that the convective eddies cannot cross. As burning at the bottom of the O shell increases the entropy (see right panel of Figure~\ref{fig:figure_10}), the stabilizing gradient in Zone~I gradually becomes weaker, and disappears altogether after $\mathord{\approx}\, 250\, \seconds$, allowing the O and Ne shell to merge into one large convection zone extending between $1.7\mathord{-}3.6\, \msolar$. The disappearance of Zone~I is mirrored by the disappearance of the entropy jump at $2.4\, \msolar$. This increase in entropy towards collapse is characteristic of the late burning stages prior to collapse, when neutrino cooling is too slow to balance thermal energy deposition due to rapid burning in the contracting shells. Zone~II also becomes narrower after $\mathord{\approx}\, 350\, \seconds$, and internal gravity waves can transport more energy across it. Different from Zone~I, this stable barrier is not completely eliminated prior to collapse, however. The right panel of Figure~\ref{fig:figure_02} shows $\omega_{\text{BV}}$ for the 1D model.
The convectively stable zones present in the 1D model are much thinner and remain unchanged over the course of the simulation. In the last $\mathord{\approx}\, 60\, \seconds$ there is a reduction in the strength of Zone~I. The contraction of the Si core after the end of the last Si shell burning phase, and deleptonisation of the Fe core leading to contraction, drives more burning (at the base of) the O shell. This raises the entropy and weakens the `buoyancy gap' between the two shells (in Zone~I). At the same time, Zone~II and Zone~III are totally undisturbed in the 1D model.
\subsection{Flow Dynamics}
\subsubsection{Growth of Density and Velocity Fluctuations}\label{sec:DensityFluctuations}
\begin{figure*}[!]
\begin{center}
\includegraphics[scale=1.0]{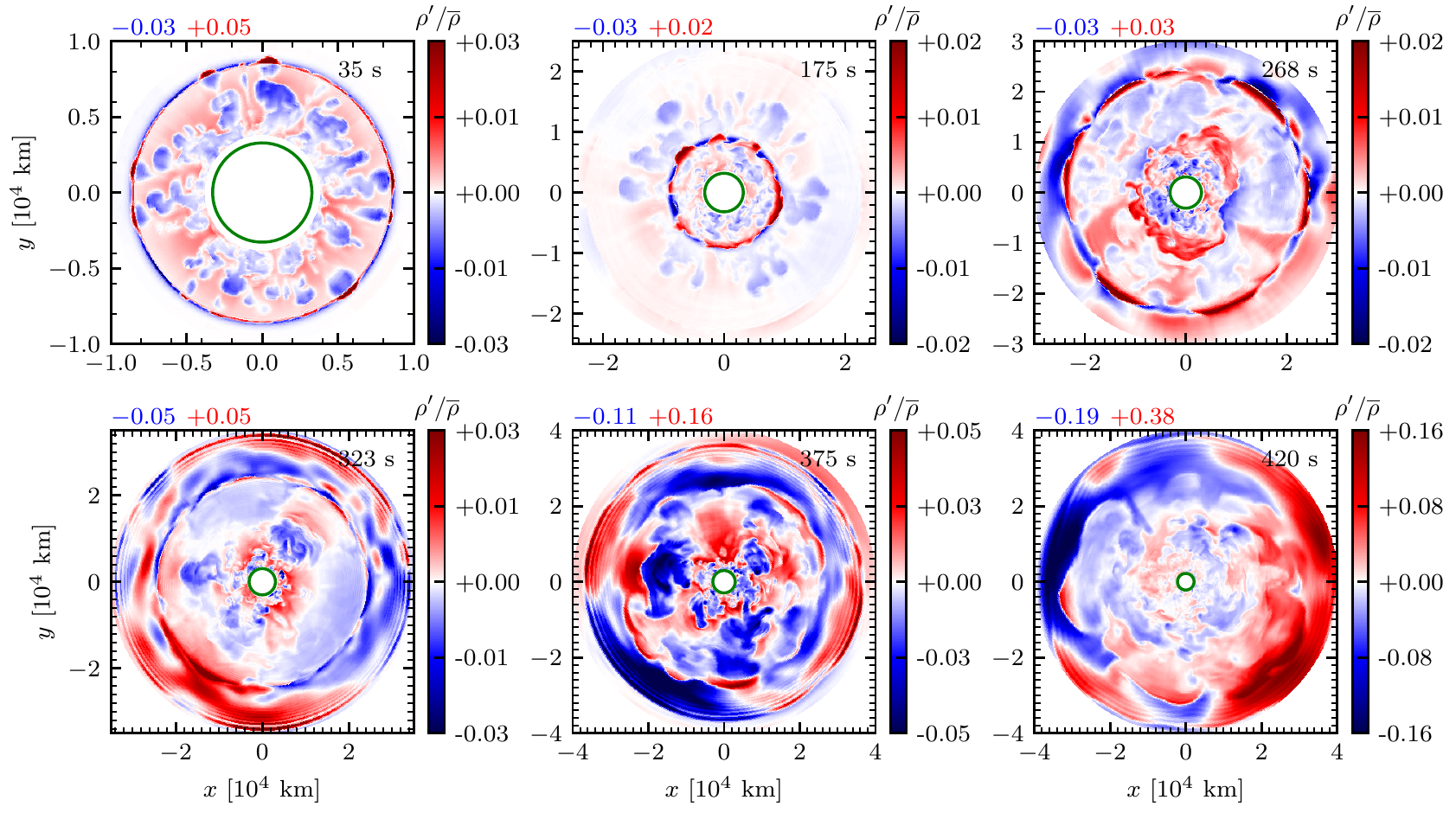}
\end{center}
\vspace*{-7mm}
\caption{Slices ($x\mathord{-}y$ plane) showing the density fluctuations $ \rho'/\overline{\rho} $ at different times. The physical size  of the region displays ranges from $10,000\, \kms$ at early time to $40,000\, \kms$ at late time. The time sequence demonstrates the transition from initially well-separated convective shells with strong density fluctuations at the boundaries from overshooting to a merging of the O and Ne shells (panels at $268\, \seconds$ and $323\, \seconds$), and eventually the C shell (panels at $375\, \seconds$ and $420\, \seconds$).  Towards the end large-scale asymmetries dominate the convective flow (see text for details). The green circle marks the inner boundary of the computational domain. Minimum and maximum values in the plane are written above the top left corner of each panel. Each panel shows data only inside a spherical region bounded by a convectively stable layer (Zone~I at $35\, \seconds$, Zone~II for $175$, $268$, $323\, \seconds$ and Zone~III for $375$ and $420\, \seconds$) for clarity of presentation. Please see the animation provided as a supplementary material (Movie~A).}
\label{fig:figure_03}
\end{figure*}
Consider the density fluctuations $\rho'$ which are defined as
\begin{equation}
    \frac{\rho' }{\overline{\rho}} 
    \coloneqq \frac{\rho(\mathbf{r}) -\overline{\rho}}{\overline{\rho}},
\end{equation}
where $\overline{\rho}$ is the average density evaluated over a spherical shell. Figure~\ref{fig:figure_03} shows the density fluctuations on 
an $x\mathord{-}y$ slice at different times. The panel at $35\, \seconds$ shows the development of plumes with density fluctuations at a level of few percent. These plumes are contained in the region below the first convectively stable layer (Zone~I) interior to $\mathord{\approx}\, 8000\, \kms$ as described in Section~\ref{subsec:conv-stab}; we refer to these as ``primary          plumes''. Primary plumes overshoot into the stably stratified layer above and are decelerated, creating ``hot spots'' in the density fluctuations as seen in both panels at $35\, \seconds$ and $175\, \seconds$. They also excite internal gravity waves which transport energy across the stable zone and thus perturb the region directly above it, creating ``secondary'' plumes. In due course the mass entrainment caused by interfacial shear instabilities driven by convection scour material off the stable interface \citep[see Figure~1 and 3 there]{strang_01}. As a result of the entrainment and mixing the stabilizing gradient in Zone~I ceases to exist around $250 \, \seconds$ (see Figure~\ref{fig:figure_02} left panel). This initiates the formation of a large convective region extending from the base of the O shell to the base of Zone~II. The eddies overshooting into Zone~II are decelerated, creating new hot spots close to $\mathord{\approx}\, 20,000\, \kms$, At this point, global asymmetries develop in the flow. The panel at $323\, \seconds$ in Figure~\ref{fig:figure_03} shows an asymmetric feature with a size of  $\mathord{\approx}\, 20,000\, \kms$, which characterizes the phase of vigorous convective activity. Close to the end of simulation (panels at $375\, \seconds$ and $420\, \seconds$) the density fluctuations in the region within $30,000\, \kms$ reach a level of $\mathord{\approx}\, 5\mathord{-}10\%$ and are again of smaller scale structure, while in the region between $\mathord{\approx}\, 30,000$ and $40,000\, \kms$ they are as large as $\mathord{\approx}\, 30\mathord{-}40\%$  with a marked dipolar asymmetry.

The cautious reader may have also noticed the sharp discontinuities in all the panels, and the associated ``waves'' in the panels from $323\, \seconds$ onward. The reason is that the fluctuation 
field renders the convective boundaries sharp; the convective 
flow rapidly decelerates due to buoyancy braking over a short 
length scale ($\mathord{\sim}\, 0.1H_{P}$, where $H_{P}$ is the 
local pressure scale height). The flow deposits energy and momentum 
and produces density fluctuations which mark the convective boundaries. The convective boundaries are stiff; or in other words have a large bulk Richardson Number \citep[see Chapter~10]{Turner_1979}, and act like a ``stretched membrane''. These boundaries reflect/transmit the internal waves generated by turbulent convection, and penetrative convection. On comparing the panels at $323\, \seconds$ and $375\, \seconds$, we see that the waves are absent outside $38,000\, \kms$, and are present in the convectively active region only. This indicates that the waves are not spurious and are certainly linked to the convective activity.
\begin{figure*}[!]
\begin{center}
    \includegraphics[width=0.95\textwidth]{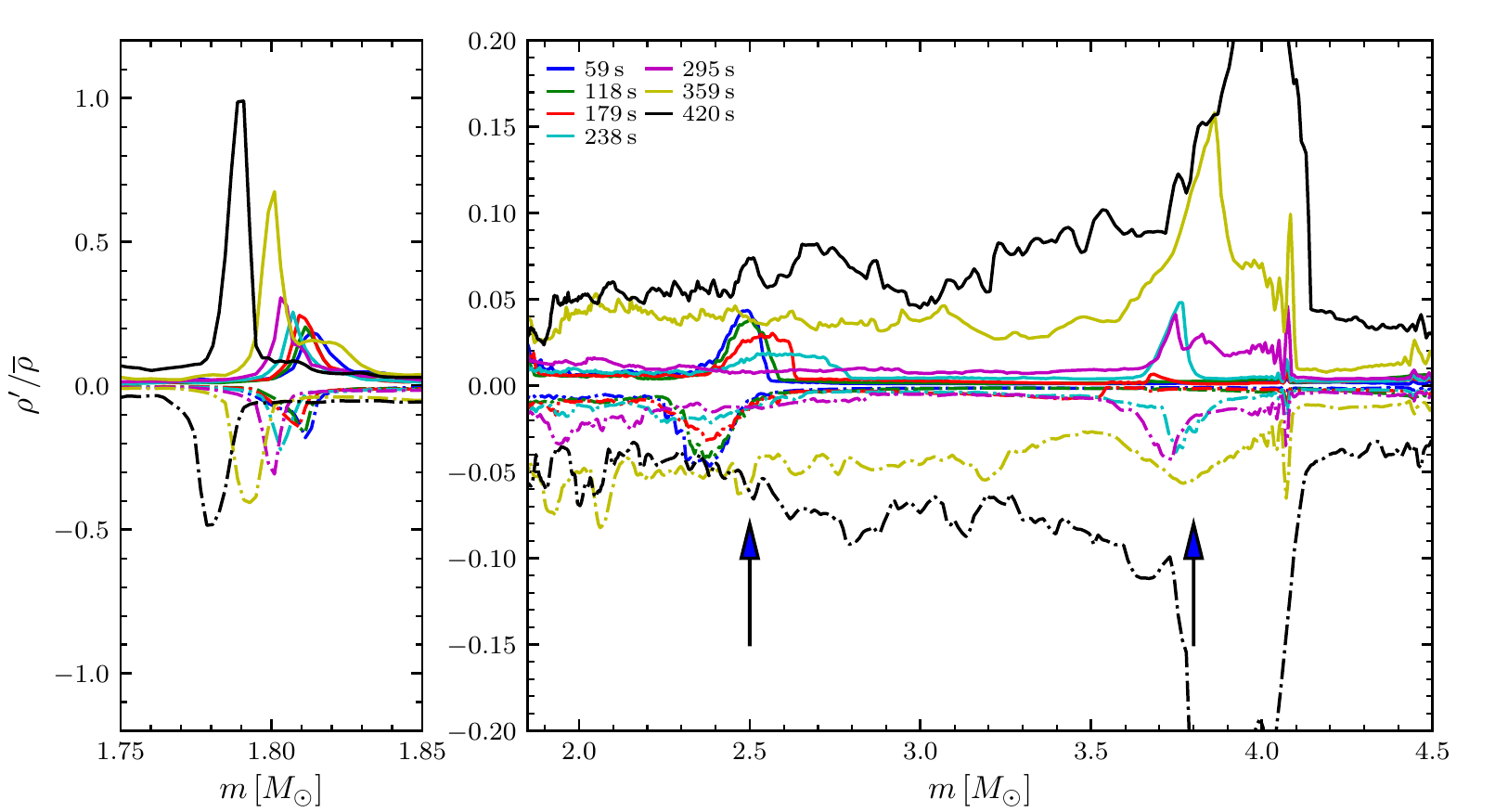}
\end{center}
\vspace*{-6mm}
    \caption{Minimum (dash-dotted line) and maximum (thick lines) values of density fluctuations $\rho'/\overline{\rho}$ as functions of mass coordinate. The arrows in the right panel mark the locations of convectively stable zones. The right panel shows the elimination of the convectively stable Zone~I (seen by large density fluctuations due to overshooting) in the process of the shell merger. In the left panel, we see that the fluctuations in the Si shell travel inwards in mass coordinate (as a consequence of slow entrainment) with time and their amplitudes grow, but they never reach the inner grid boundary at $1.75\, \msolar$. Blue arrows (right panel) mark the positions of Zone~I and Zone~II.}
    \label{fig:figure_04}
\end{figure*}
Figure~\ref{fig:figure_04} provides another perspective on the flow dynamics by showing the minimum and maximum of the fractional density fluctuations on spherical mass shells as functions of mass coordinate. One clearly sees (right panel) how the density fluctuations at shell interfaces initially decrease in magnitude and spread out towards larger $m$ as entrainment whittles down the stabilizing entropy gradients at the shell interfaces. After the shells have merged, the density fluctuations grow considerably, especially in the outer part of the convective region. Figure~\ref{fig:figure_04} (left) shows that the inner convective boundary moves towards lower $m$ by entrainment; here the density fluctuations in the boundary layer actually become stronger with time as the overall violence of convection increases. We also point out that the density fluctuations never touch the inner grid boundary, which confirms the proper choice of its location at $1.75\, \msolar$.
\begin{figure*}[!]
\begin{center}
\includegraphics[scale=1.0]{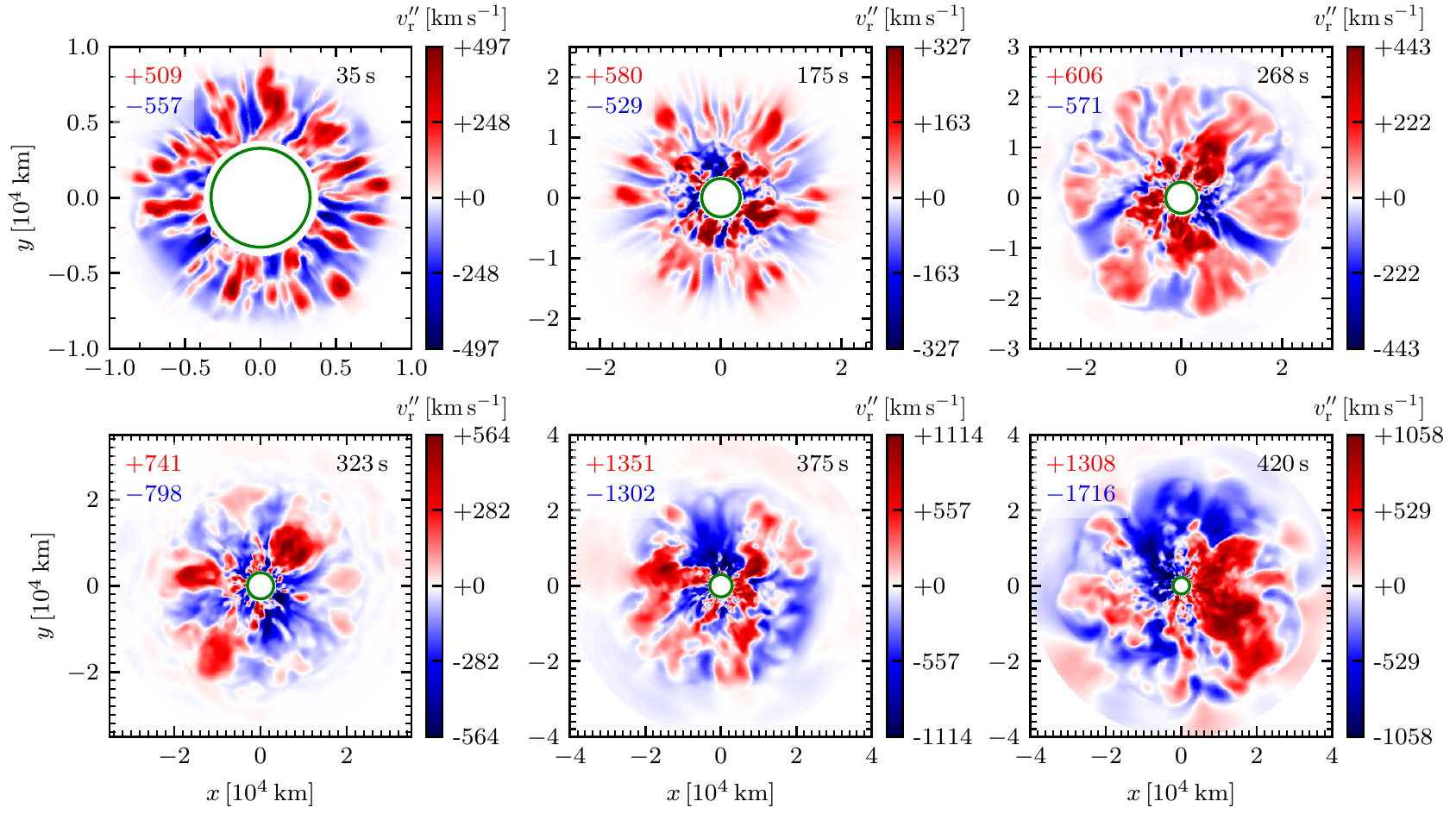}
\end{center}
\vspace*{-6mm}
\caption{Slices in the $x$-$y$ plane showing radial velocity fluctuations $v''_{\text{r}}$ (in $\kmps$) at different times. The panels $35\, \seconds$ till $268\, \seconds$ show the initially slow but steady build-up of the convective velocities. After the merger of the O and Ne shells, convection becomes considerably more violent, and large-scale flow patterns develop (see the text for details). The green circle marks the inner boundary of the computational domain. Minimum and maximum values in the plane are specified in the top left corner of each panel. Each panel shows data only inside a spherical region bounded by a convectively stable layer (Zone~I at $35\, \seconds$, Zone~II for $175$, $268$, $323\, \seconds$ and Zone~III for $375$ and $420\, \seconds$) for clarity of presentation. Please see the animation provided as a supplementary material (Movie~A).}
\label{fig:figure_05}
\end{figure*}
We next consider the turbulent velocity fluctuations.
Because the model is initially non-rotating, we do not include any
non-radial components in the mean flow and decompose the velocity 
field as (see Appendix~\ref{sec:favre-avg})
\begin{equation}
\mathbf{v}(\mathbf{r}) = \tilde{v}_{r}(r)\mathbf{e}_{\mathbf{r}}+v_{r}''(\mathbf{r})\mathbf{e}_{\mathbf{r}} +
v_{\theta}\mathbf{e}_{\theta}+v_{\phi}\mathbf{e}_{\phi},
\end{equation}
where $\tilde{v}_{r}$ is the Favre-averaged radial velocity. 
The fluctuating component $v_{r}''$ of  the radial velocity is therefore given by 
\begin{equation}
    v''_{r}(\mathbf{r}) \coloneqq v_{r}(\mathbf{r}) -\tilde{v}_{r}. \label{eqn:vel-fluc}
\end{equation}
Figure~\ref{fig:figure_05} shows the radial velocity fluctuations corresponding to the snapshots shown in Figure~\ref{fig:figure_03}. The panel at $35\, \seconds$ shows well-developed plumes extending from the O burning
region at $\mathord{\approx}\, 3,000\, \kms$ to the base of Zone~I at $8,000\, \kms$. The panel at $175\, \seconds$ shows the secondary plumes extending from $\mathord{\approx}\, 8,000\, \kms$ to $\mathord{\approx}\, 20,000\, \kms$. The primary and secondary plumes are physically separated from each other by the Zone~I. Also, the secondary plumes have lower velocities (by a factor of $2\mathord{-}3$) compared to the primary plumes. These plumes are driven by the active Ne shell and develop later because of a lower Brunt-V{\" a}is{\" a}l{\" a} frequency. The velocity difference is also apparent in panel at $268\, \seconds$ which is the stage when the stable Zone~I has already disappeared.The panel at $323\, \seconds$ shows emergence of a large-scale flow spanning the entire region from base of the oxygen burning layer to the base of Zone~II. At this stage, convection becomes very vigorous and the magnitude of velocity fluctuations increases from $\mathord{\approx}\, 800\, \kmps$ to more than $1,600\, \kmps$ (panels at $375\, \seconds$ and $420\, \seconds$) within the next $120\, \seconds$. The panel at $420\, \seconds$ shows the presence of a large-scale dipolar asymmetry in the flow. 
Convective downdrafts in the innermost region reach velocities in excess of $1,600\, \kmps$ aided by rapid contraction of the Fe/Si core.
\subsubsection{Convection Length Scale}\label{sec:corr_len}
\begin{figure*}[!]
\begin{center}
\includegraphics[scale=1.0]{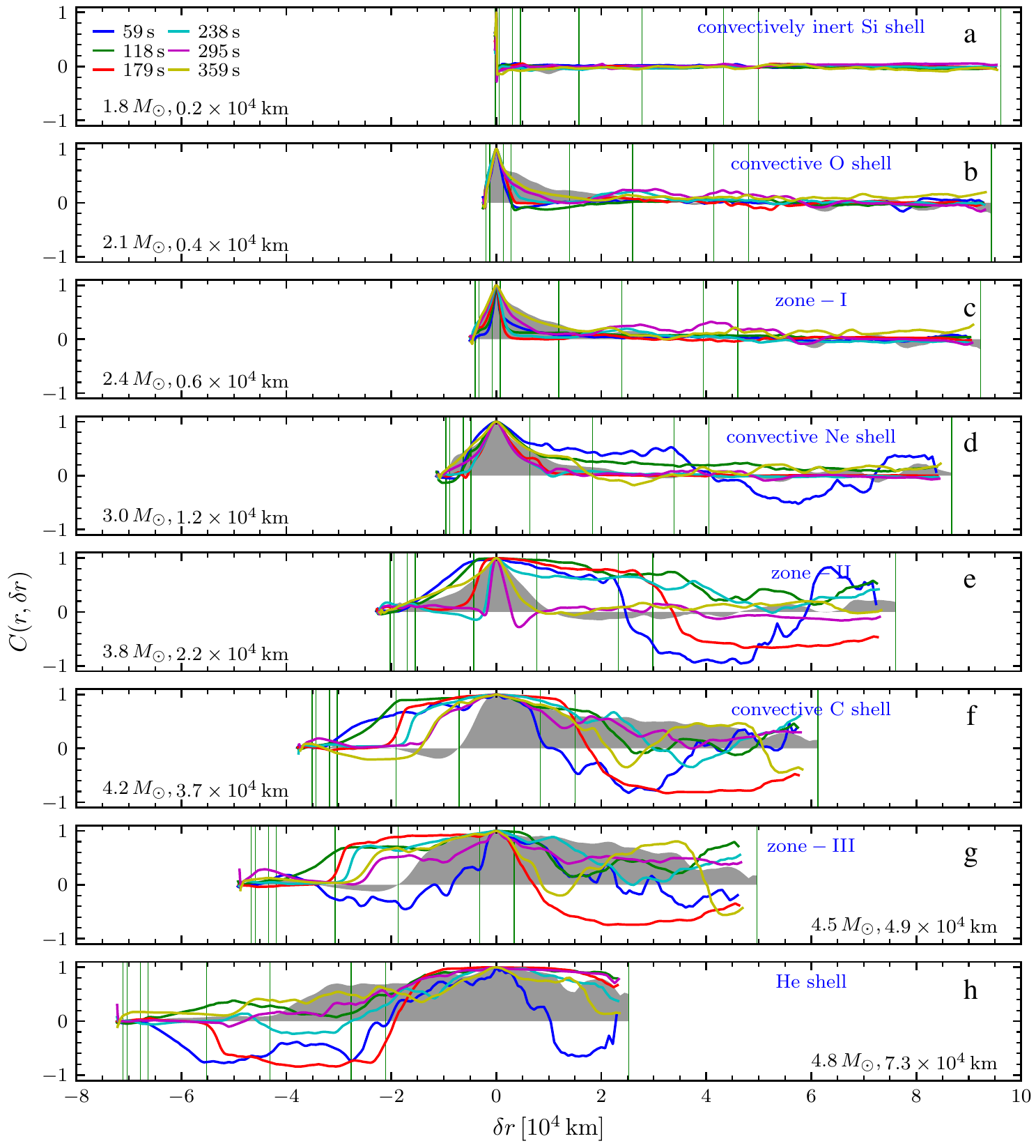}
\end{center}
\vspace{-6mm}
\caption{Velocity correlation functions (defined by Equation~\eqref{eqn:vcorr}) near the centers of layers a$-$h (listed in Section~\ref{sec:corr_len}) at various times. The mass and radius coordinates of these points are written in the corresponding panels. Vertical green lines mark the boundaries of these layers. The correlation function at $420\,\seconds$ is shown as filled grey curve in each panel. Placing the panels~d and h together (aligning them along the green lines) shows two well developed convective layers with a contact somewhere inside Zone~II. See the text for details.}\label{fig:figure_06}
\end{figure*}
The initial seed perturbations are applied 
as random cell-by-cell variations and hence on a very short length scale. 
In the absence of a buoyancy barrier convective flow naturally grows to the largest possible unimpeded length scale.
The longitudinal two-point correlation function gives a qualitative measure of the size of the convective region. It is defined as (Equation~15 of \citealp{mueller_16c})
\begin{equation}
    C(r,\delta r) \coloneqq \frac{\overline{{v'}_{r}(r,\theta,\phi){v'}_{r}(r+\delta r,\theta,\phi)}}{\left\{\overline{{v'}_r^2(r,\theta,\phi)}\quad \overline{{v'}_r^2(r+\delta r,\theta,\phi)}\right\}^{1/2}}, \label{eqn:vcorr}
\end{equation}
where we have used the Reynolds-averaged velocities (see Appendix~\ref{sec:favre-avg}). The correlation length $\Lambda_{\text{corr}}$ at a radius $r$ is defined by integrating the correlation function 
\begin{equation}
    \Lambda_{\text{corr}}(r) \coloneqq \int\limits_{r_{-}}^{r_{+}}C(r,\delta r)\, \ud r', \quad \delta r = r'-r,
\end{equation}
between $r_{-}$ and $r_{+}$, which are usually taken to be $-\infty$ and $+\infty$ respectively. \citet[Appendix~B]{meakin_07_b} define these points such that $C(r,r_{\pm}-r)\, \mathord{\geq}\, 0.5$. In the present case, estimating the correlation length using the equation above is non-trivial because of the presence of multiple convectively stable zones. For this analysis we divide the simulation domain into multiple layers according to their initial convective stability and the burning processes:
\begin{enumerate}
 \setlength\itemsep{-0.005em}
    \item[\textit{a.}]\, $1.7\mathord{-}1.8\, \msolar$, convectively inert Si shell,
    \item[\textit{b.}]\, $1.8\mathord{-}2.4\, \msolar$, convective O shell,
    \item[\textit{c.}]\, $2.4\mathord{-}2.5\, \msolar$, Zone~I,
    \item[\textit{d.}]\, $2.5\mathord{-}3.6\, \msolar$, convective Ne shell,
    \item[\textit{e.}]\, $3.6\mathord{-}4.1\, \msolar$, Zone~II,
    \item[\textit{f.}]\, $4.1\mathord{-}4.4\, \msolar$, convective C shell,
    \item[\textit{g.}]\, $4.4\mathord{-}4.5\, \msolar$, Zone~III,
    \item[\textit{h.}]\, $4.5\mathord{-}5.0\, \msolar$, He shell.
\end{enumerate}\label{list:layers}
Figure~\ref{fig:figure_06} shows the correlation function evaluated at roughly the middle points of these layers. We have marked the boundaries of these layers by vertical lines. In panel~a the correlation function $C(r,\delta r)$ drops rapidly from unity with $\delta r$, resulting in $\Lambda_{\text{corr}}\,\mathord{<}\, 1,000\, \kms$ at all times ($\Lambda_{\text{corr}}/r\,\mathord{\ll}\, 1$). Thus the Si layer underneath the O burning zone stays non-convective and dynamically disconnected from the overlying shells during the entire course of the evolution. The correlation function has a finite width ($\Lambda_{\text{corr}}\, \mathord{\lesssim}\, r$) in panels~b, c, and d such that its value approaches zero for $\delta r\, \mathord{\approx}\, 10,000\, \kms$. The correlation function approaches a symmetrical shape as we traverse the O and Ne shells. 
If the correlation function is evaluated at the centre of a convective region then it will have a symmetrical shape, which will change as we go to the bottom/top of the region. As one moves from the Ne shell into Zone~II the function again becomes asymmetric, which implies that after $\mathord{\approx}\, 180\, \seconds$ there is a well defined convective region centered somewhere between $3.0\mathord{-}3.8\, \msolar$. As we traverse Zone~III and go into the He shell, the correlation length becomes large ($\Lambda_{\text{corr}}\mathord{\sim}r$) and the function approaches a symmetrical form. This suggests that after $\mathord{\approx}\, 300\, \seconds$ there is a well defined convection layer centered somewhere between $4.5\mathord{-}4.8\, \msolar$. The final state of the simulated shell has two well defined convective regions with a contact somewhere inside Zone~II. 
\subsection{Thermodynamic Evolution}\label{sec:TE}
The continuous deposition of thermal energy by nuclear burning powers convection. In the following sections we compare nuclear energy production and entropy evolution in the 1D and 3D models.
\subsubsection{Nuclear Energy Generation: 1D vs. 3D}\label{sec:NucGen3D}
\begin{figure*}[!]
\vspace{8mm}
\begin{center}
\includegraphics[width=\linewidth]{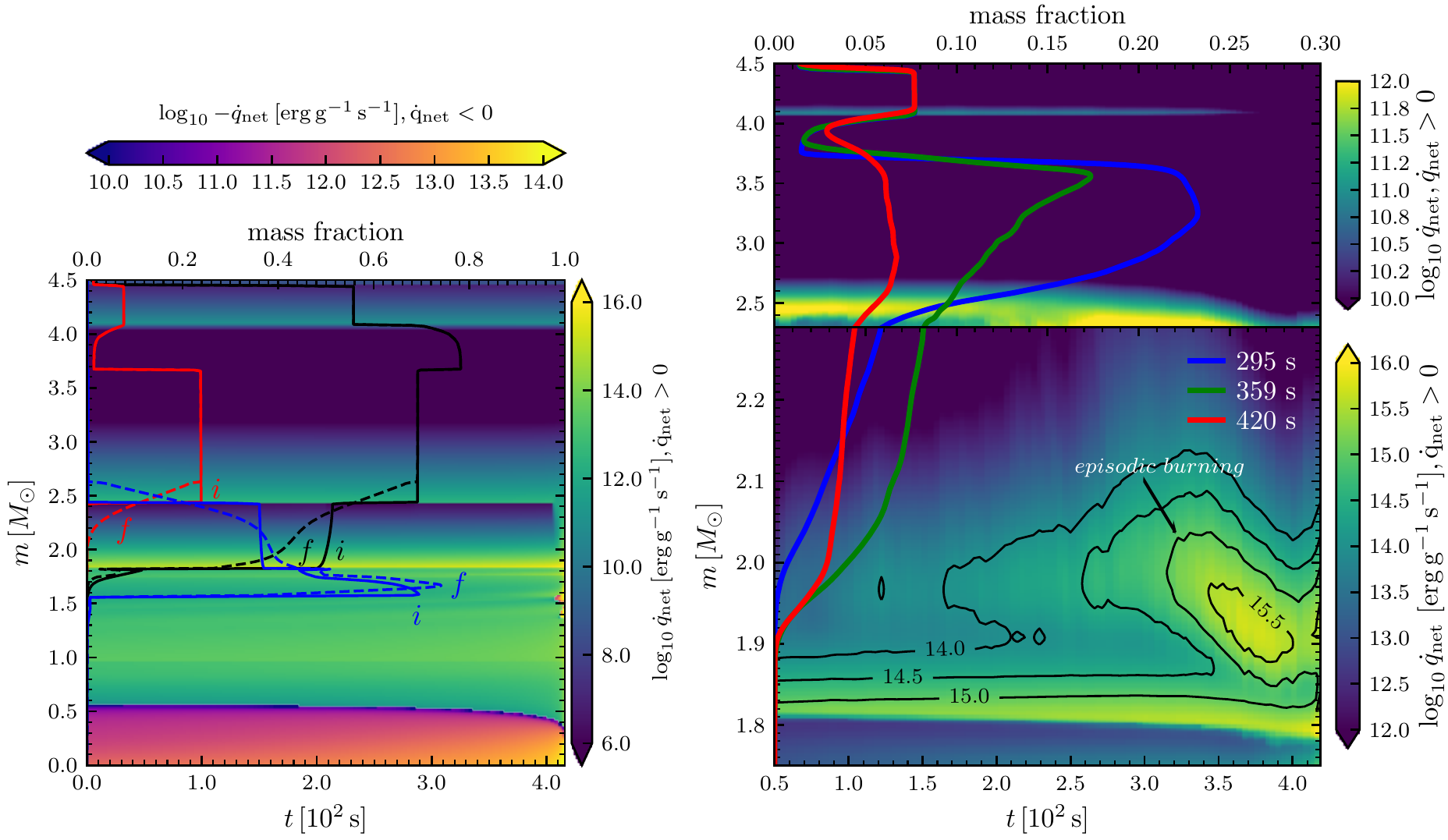}
\end{center}
\vspace{-6mm}
\caption{Left panel: Space-time plot showing net energy generation ($\dot{q}_{\text{nuc}}\mathord{-}\dot{q}_{\nu}$ ,color-coded) rate for the 1D model. The solid curves (\textit{i}) show initial mass fraction profiles and the dashed curves (\textit{f}) show the final mass fraction profile for O (black), Ne (red) and Si (blue) with the corresponding scale on the top of the panel. Burning happens at the base of the O shell close to $1.8\, \msolar$. Neutrino cooling is dominant in the central part of the Fe/Si core (top colorbar). The burning rate is practically constant in time. The right panel shows the color-coded net energy generation rate for the 3D model. The upper panel shows the region inside $2.3\mathord{-}4.5\, \msolar$. The energy generation is relatively constant in time except some changes close to the end resulting from Ne depletion. The lower panel shows a zoom of the region inside $1.8\mathord{-}2.3\, \msolar$. The energy production here is dominated by O burning at base of the O shell (see the contour at $\log_{10}\dot{q}_{\rm net}\,\mathord{=}\,15$) till $\mathord{\approx}\, 300\,\seconds$. Material entrained from the Ne shell burns on its way down forming a secondary burning layer separated from the O burning zone. At $\mathord{\approx}\, 300\, \seconds$ there is a rapid increase in the spatial extent and energy production rate inside the secondary layer. This marks the \textit{episodic burning} phase; a significant amount of Ne entering the O shell is ignited and consumed over a short duration. The colored lines represent the Ne mass fraction at different times(scale at the top of the panel).}\label{fig:figure_07}
\end{figure*}
\begin{figure*}[!]
    \begin{center}
    \includegraphics[width=\linewidth]{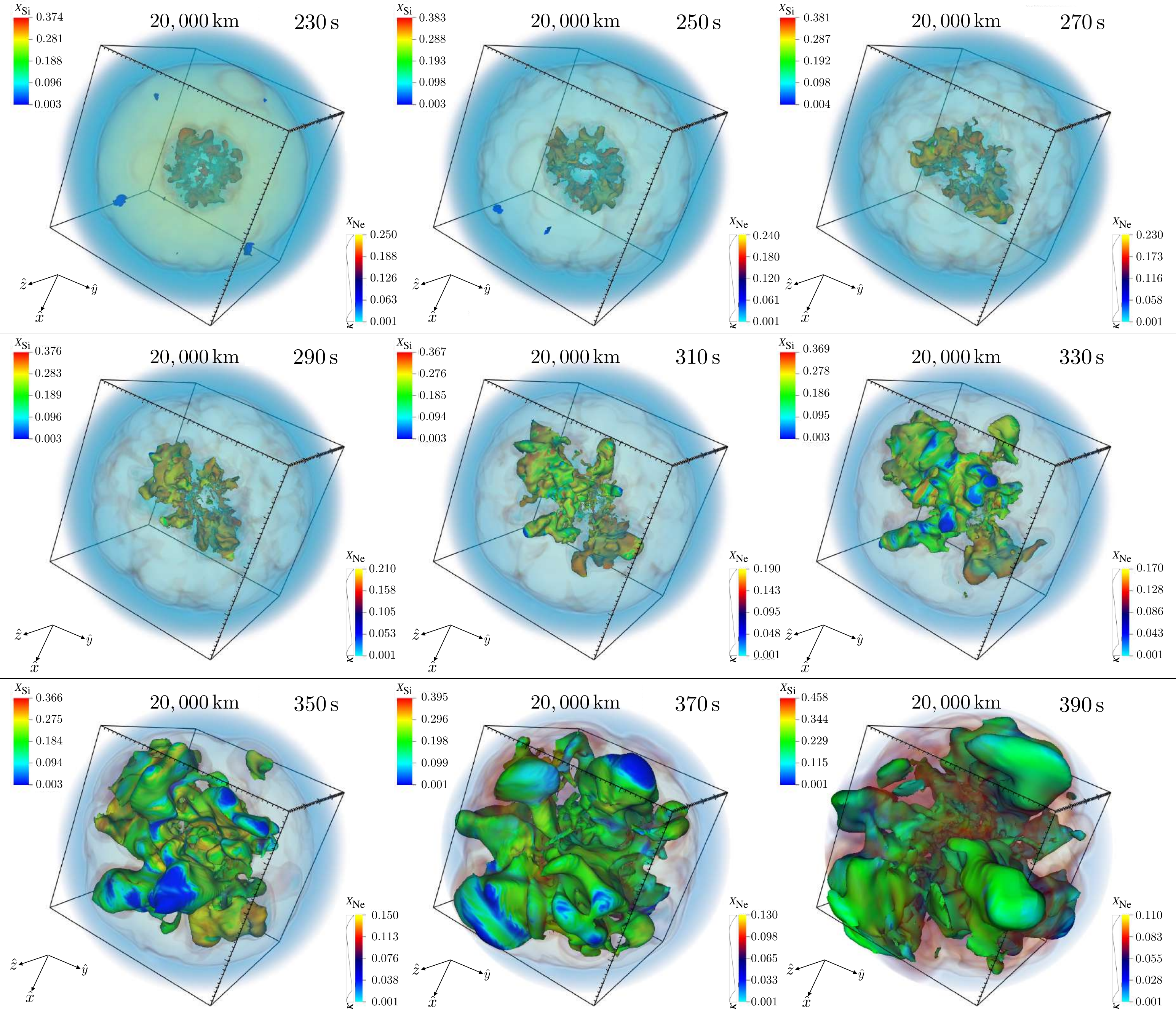}
    \end{center}
    \vspace*{-4mm}
    \caption{Sequence of 3D volume renderings following the shell merger in progress. The Ne mass fraction is volume rendered and appears as nearly a transparent ``cloud'' (with the opacity function $\kappa$ shown in the lower right corner of each panel). The inner opaque bubble with a convoluted morphology represents a color plot of the Si mass fraction (the color coding of latter is given in the upper left corner of each panel) on a radial velocity isosurface (at $250\, \kmps$).
    The choice of radial velocity for the isosurface is motivated by tracking the average value of Si mass fraction. The box is $2\mathord{\times} 20,000\, \kms$ across, which is the extent of the main Ne shell (see Figure~\ref{fig:figure_01}). The progression of states (from left to right and top to bottom) shows the violent O-Ne shell merger in progress.}
    \label{fig:figure_08}
\end{figure*}
\begin{figure*}[!]
\begin{center}
\includegraphics[scale=1.0]{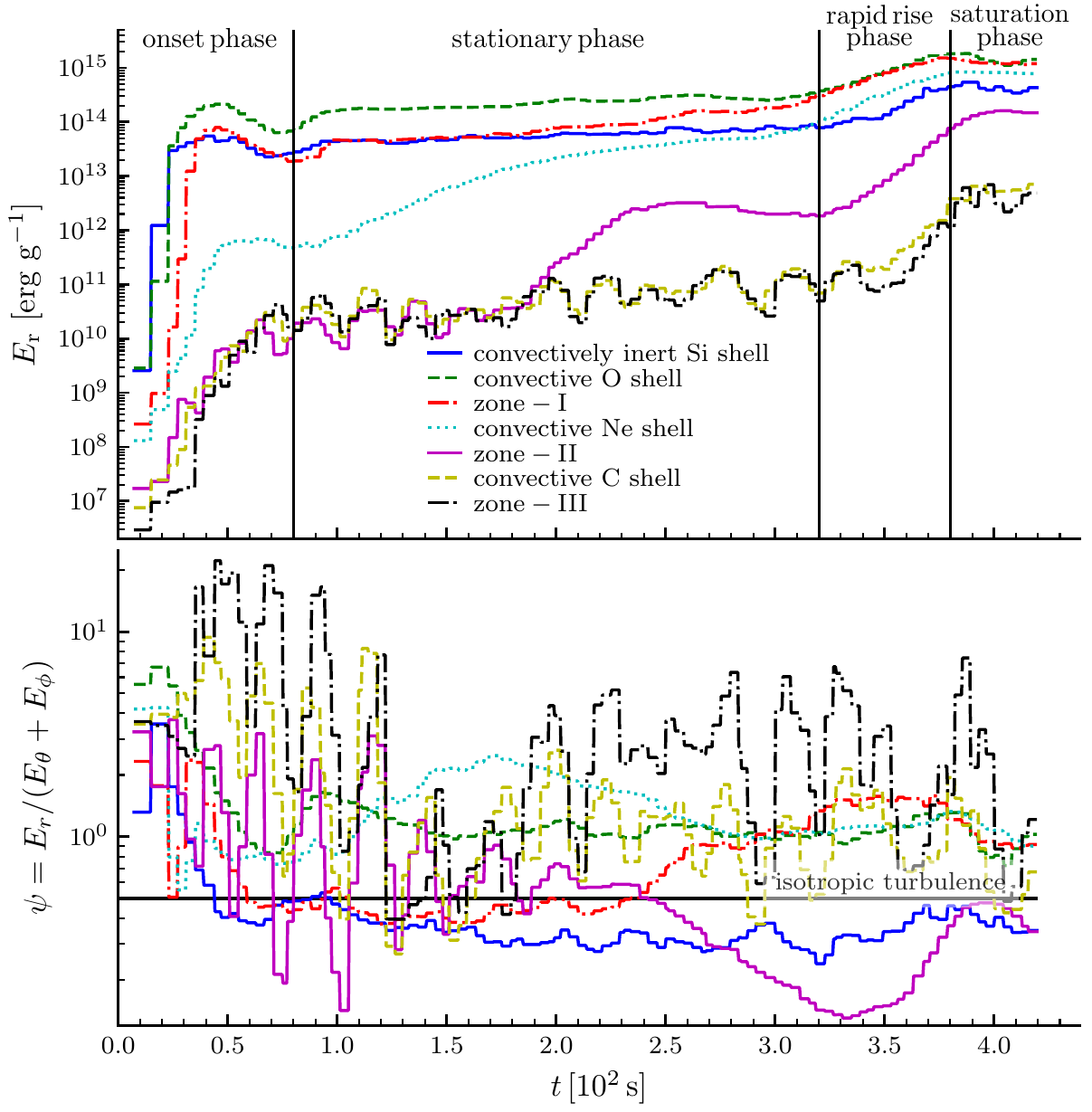}
\end{center}
\vspace*{-6mm}
\caption{Upper panel: Specific kinetic energies in the radial velocity fluctuations as functions of time for the different layers. The evolution can be split into four phases: a) onset of convection, b) stationary phase, c) rapid rise, and d) saturation phase. The entire region between the inert Si shell and the Ne shell attains similar specific kinetic energy after the shell merger. Lower panel: Ratio of kinetic energy stored in the radial fluctuations ($E_{\text{r}}$) and the transverse flow ($E_{\theta}+E_{\phi}$) as a function of time. The horizontal line marks the value (1/2) when the kinetic energy is equally partitioned between all the components as in case of isotropic turbulence, showing that the flow is primarily dominated by large-scale radial motions due to convection.}
\label{fig:figure_09}
\end{figure*}
Figure~\ref{fig:figure_07} (left panel) shows the net energy generation rate in the 1D model as a function of time. In the region inside $0.6\, \msolar$ (at higher density and temperature inside the Fe core) the neutrino cooling rate dominates over the nuclear energy generation rate leading to core contraction. In the Si layer energy deposition rate by nuclear burning of Si to Fe exceeds the energy loss rate due to neutrino cooling.  Most of the nuclear energy generation happens at the base of the O shell between $1.8\mathord{-}2.1\, \msolar$. In addition, there are contributions from Ne shell burning, C shell burning, and He shell burning (not shown in the figure). Note that the C shell burning and the He shell burning are not relevant for the dynamics. Also, the total energy generation in the He shell is negligible.  No
energy from it enters or leaves the simulation
domain, as its main purpose is to provide `boundary pressure'.

Figure~\ref{fig:figure_07} (right panel) shows the net energy generation rate ($|\dot{q}_{\text{nuc}}|\, \mathord{\gg}|\, \dot{q}_{\nu}|$ everywhere) in the 3D model.
The evolution between $2.3\mathord{-}4.5\, \msolar$ (upper half) does not show a lot of deviation from its initial behaviour, except that close to $\mathord{\approx}\, 300\, \seconds$ the Ne burning shell moves inwards in mass (in a Lagrangian sense). The lower half shows a zoom of the region between $1.8\mathord{-}2.3\, \msolar$. Most of the energy production happens at the base of the O shell. Around $100\, \seconds$ another burning layer develops above the O burning zone but below the Ne shell. Ne entrained from the Ne shell (see colored lines representing the Ne mass fraction at different time) burns on its way to the base of the O shell. Both the extent of this layer and intensity of burning increases with time, and around $300\, \seconds$ there is a short but intense Ne burning episode. This phenomenon is marked as \textit{episodic burning}, which is reminiscent of \textit{hot-spot burning} \footnote{Note that the term ``hot-spot burning'' is being used here in a figurative sense. 
It is meant to describe a place of vigorous nuclear burning activity; large-scale mixing of Ne deep in to the O shell and its subsequent burning at the higher 
ambient density and temperature. We think that our usage of the term ''hot-spot burning'' confirms with 
the phenomenon described by \citet[see their Figure~3]{bazan_94}. Quoting \citet{bazan_98},
``\textit{Significant mixing beyond convective boundaries determined by
mixing-length theory brings fuel ($^{12}\rm C$) into the 
convective region, causing hot spots of nuclear burning}.''} seen by \citet[see Fig.~3]{bazan_94} in their 2D study of O burning in a $20\, \msolar$ progenitor \citep{arnett_94}. 

Figure~\ref{fig:figure_08} shows the resulting shell merger marked by Si rich outflow rapidly moving into the Ne shell. The isosurfaces shown track the points corresponding to $250\, \kmps$ radial velocity. The Si-rich material moves rapidly outwards carving out an elongated cavity in the enclosing Ne shell. The velocity isosurface expands from $\mathord{\approx}\, 10,000\, \kms$ to more than $20,000\, \kms$ in a short span of
$\mathord{\approx}\, 160\, \seconds$.
The episodic nuclear burning peaks at around $350\, \seconds$ (as shown in Figure~\ref{fig:figure_07} by $\log \dot{q}$ contours) and triggers/powers a phase of violent convective activity. A comparison between the Ne abundance at $360\, \seconds$ and $420\, \seconds$ shows that the total amount of Ne decreases by $\Delta M_{\text{Ne}}\mathord{\approx}\, 0.2\, \msolar$ ($\mathord{\approx}$ 70 percent) because of the episodic burning. 

A rough estimate of the energy produced can be obtained by considered the principle energy producing reactions in Ne burning which effectively convert two $^{20}\text{Ne}$ nuclei to $^{16}\text{O}$ and $^{24}\text{Mg}$ nuclei \citep[Equation~5.108]{iliadis_2015}
\begin{equation}
^{20}\text{Ne} + ^{20}\text{Ne} \to ^{16}\text{O} + ^{24}\text{Mg},
\end{equation}
with the average energy production (per $40$ nucleons) of $\overline{Q}_{\text{Ne}}\mathord{\approx}\, 6.2\, \text{MeV}$ ($\mathord{\equiv}\, 1.5\mathord{\times}10^{17}\, \ergsg$) near $T\, \mathord{\approx}\, 1.5\, \text{GK}$ \citep[Equation~5.109]{iliadis_2015}. This implies a total release of $\overline{Q}_{\text{Ne}}\Delta M_{\text{Ne}}\, \mathord{\approx}\, 6.0\mathord{\times}10^{49}\, \text{erg}$ of thermal energy inside $\mathord{\approx}\, 2.5\, \msolar$ of stellar plasma within $\mathord{\approx}\, 100\, \seconds$, which further translates into $\mathord{\approx}\, 4\mathord{\times}10^{15}\, \ergsg$ in each component ($E_{\text{r}}$, $E_{\theta}$ and $E_{\phi}$) of kinetic energy. 

The upper panel of Figure~\ref{fig:figure_09} shows the growth of the specific kinetic energy (per unit mass) in the fluctuating component in all of the layers, individually defined as
\begin{equation}
E_{\text{r}}(i) \coloneqq \frac{1}{2\delta m_{i}}\int\limits_{m_{i,-}}^{m_{i,+}}\rho v''^2_{r}\, \ud m,
\end{equation}
where $m_{i,-},m_{i,+},\delta m_{i}$ are the inner boundary, outer boundary and mass, respectively, of the $i$th layer (defined in Section~\ref{sec:corr_len}). Kinetic energy starts building up with the \textit{onset} of convection and reaches a plateau phase by $\mathord{\approx}\, 70\mathord{-}80\, \seconds$ except for the Ne shell. The \textit{stationary phase} lasts till $\mathord{\approx}\, 320\, \seconds$, when episodic Ne burning 
begins. In the next $\mathord{\approx}\, 20\mathord{-}30\, \seconds$ there is a \textit{rapid rise} in kinetic energy by a factor of $\mathord{\sim}10$. In the end the growth \textit{saturates}, and the entire merged shell has uniform specific kinetic energy. The lower panel of Figure~\ref{fig:figure_09} shows the ratio $\psi$ of kinetic energy in the fluctuating radial component and the transverse components. The transverse components of the specific kinetic energy are defined as
\begin{equation}
E_{\text{a}}(i) \coloneqq \frac{1}{2\delta m_{i}}\int\limits_{m_{i,-}}^{m_{i,+}}\rho v^2_{\text{a}}\, \ud m,\quad \text{a} \in\{\theta,\phi\}.
\end{equation}
The ratio fluctuates with time and hardly stays close to the equipartition value ($E_{r}=E_{\theta}=E_{\phi}$) for the convectively active shells. For the stable Zone~I the value deviates away from $1/2$ when the shell is finally eroded away. Interestingly, transverse motions dominate over radial motion in Zone~II after $\mathord{\approx}\, 250\seconds$.
\subsubsection{Entropy: 1D vs 3D}
\begin{figure*}[!]
\begin{center}
\includegraphics[scale=1.0]{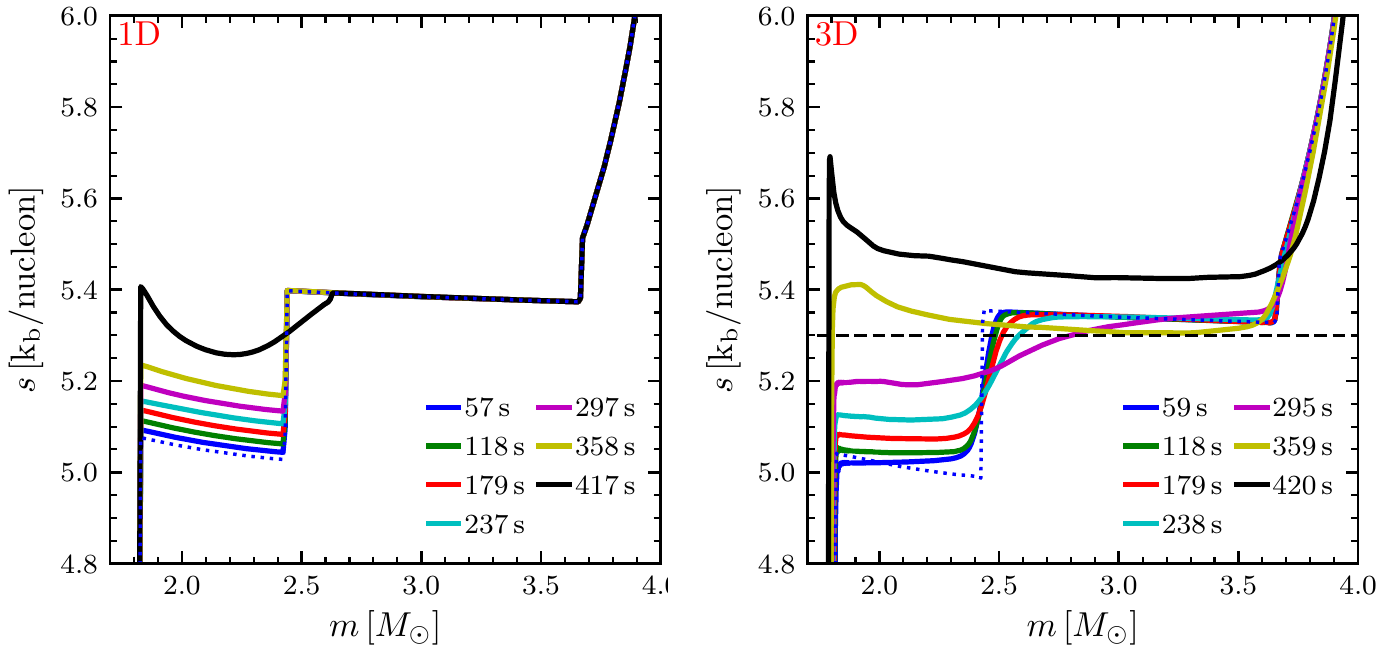}
\end{center}
\vspace*{-6mm}
\caption{Entropy evolution in the 1D model (left panel) and the 3D model (right panel). The dotted blue curve represents the initial ($t=0\, \seconds$) entropy profile in each of the panels. In the 1D model, most of the change is confined to the region between $1.8\mathord{-}2.6\, \msolar$. Energy deposited by O burning causes a gradual increase in entropy until $\mathord{\approx}\, 360\, \seconds$. At the end, the sharp jump in entropy at $\mathord{\approx}\, 2.4\, \msolar$ disappears. In the 3D model efficient heat transport quickly levels the entropy gradient inside $1.8\mathord{-}2.5\, \msolar$. In the last $\mathord{\approx}\, 120\, \seconds$ rapid energy deposition by episodic Ne burning leads to a negative entropy gradient extending from base of the O shell to the base of Zone~II. Note: The dashed horizontal line in the right panel is to guide the eye to the inversion of entropy gradient, which happens between $295\, \seconds$ and $360\, \seconds$.}\label{fig:figure_10}
\end{figure*}
Regions inside a star tend
to become convective when radiation and neutrinos are unable to carry away the thermal energy deposited by nuclear burning. In a stably stratified region the entropy gradient is positive (entropy increases radially outwards). Figure~\ref{fig:figure_10} shows the entropy profiles for the 1D model (left panel) and the 3D model (right panel). The differences between the entropy profiles of the 1D and 3D models are conspicuous. The negative entropy gradient at the base of the O layer facilitates convection.
 
In case of the 1D model, the entropy in the region inside $1.8\mathord{-}2.4\, \msolar$ increases gradually until the first $360\, \seconds$ ($\Delta s\, \mathord{\approx}\, 0.1\, \kbpnuc$), keeping the gradient unchanged. In the last $60\rm \, \seconds$, the inner Fe/Si core experiences 
significant contraction ($\approx$50\%). The associated density 
and temperature increment leads to a rapid increase in the burning
rate of oxygen as well as the neon (1D mixing). This raises the overall 
entropy in the oxygen shell and weakens the `buoyancy gap' between the O and Ne shells. The softening of convective boundary \footnote{Note: In \kepler, convective boundaries are 'softened' by one 
formal overshoot zone with $1/10$ mixing efficiency as 
semiconvection.} leads to more efficient mixing.
\begin{figure}
    \centering
    \includegraphics{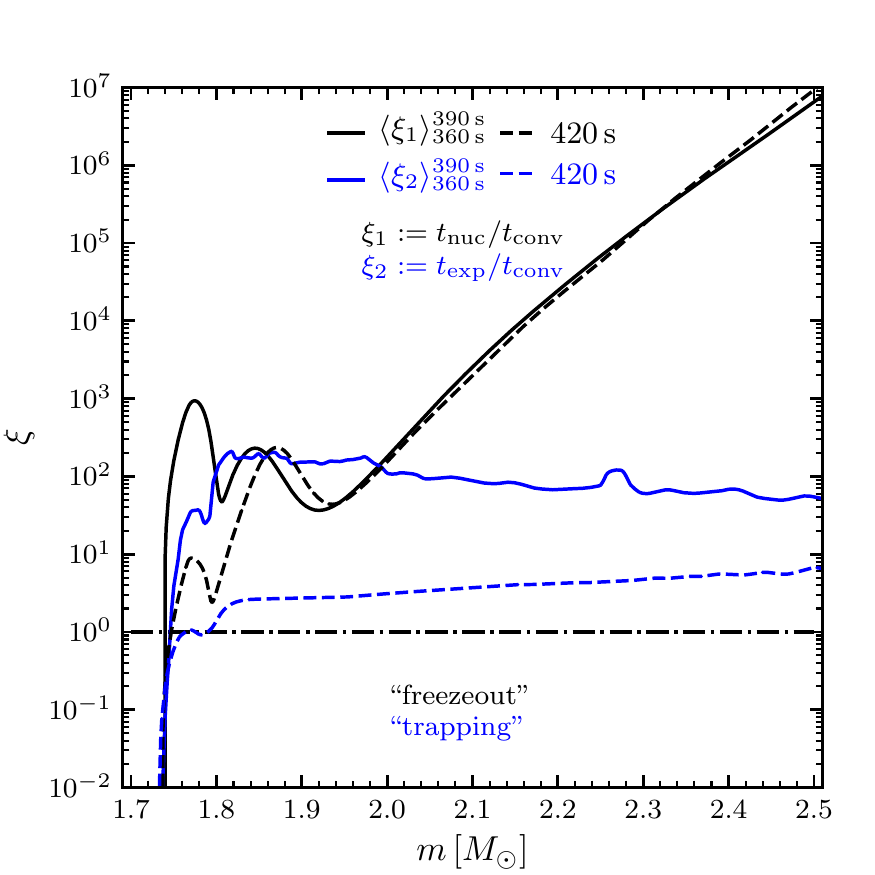}
    \caption{A comparison between nuclear ($t_{\rm nuc}$), convective ($t_{\rm conv}$), and expansion ($t_{\rm exp}$) timescales during the last $60\, \rm \seconds$ leading to collapse. Thick solid curves show the average value of $\xi_1\,\mathord{\coloneqq}\,t_{\rm nuc}/t_{\rm conv}$ and $\xi_2\,\mathord{\coloneqq}\,t_{\rm exp}/t_{\rm conv}$ over $t\in [360,390]\, \seconds$, and the thick dashed curves show the value of $\xi_1$ and $\xi_2$ just before the onset of collapse. The ratio of nuclear timescale and the convective turnover timescale $\xi_1\mathord{\gg}1$ everywhere, with the exception of the inner boundary. Close to the onset of collapse $\xi_2\mathord{\sim}1$ in the region inside $1.8\,\rm \msolar$, pointing to a rapid decline in the efficiency of convective heat transfer. This effective \emph{trapping} of the convective flow produces a peak in entropy (during the last $5\, \rm \seconds$ entropy changes by $\mathord{\approx}\,0.2\, \kbpnuc$ in the innermost region) at the base of the O burning shell.}
    \label{fig:figure_11}
\end{figure}
In case of the 3D model the entropy evolution is remarkably different. The overall entropy increase is mostly due to the heat deposited
by oxygen burning and hot-spot burning. In the first $360\, \rm \seconds$,
the average entropy for 
the matter inside $2.5\,\mathord{-}\,3.7\, \rm \msolar$ decreases by $\Delta s_{2.5-3.7}\,\mathord{\approx}\, 0.03\, \kbpnuc$. This high-entropy material 
is carried inwards by convective downdrafts. During the same time, the average entropy for 
the matter inside $1.8\,\mathord{-}\,2.5\, \rm \msolar$ increases by $\Delta s_{1.8-2.5}\,\mathord{\approx}\, 0.54\, \kbpnuc$. The ratio of $\Delta s_{1.8-2.5}$ and $\Delta s_{2.5-3.7}$ is $\mathord{>}\,10$, which clearly shows that the entropy gain of the inner region cannot be explained as resulting from entrainment of high-entropy material and must be a result of nuclear burning. The entropy profile inside $1.8\mathord{-}2.4\, \msolar$ is levelled in the first $60\, \seconds$, but the additional Ne burning preserves the small negative entropy gradient between $2.4\mathord{-}3.6\, \msolar$. At the same time the steep entropy barrier close to $2.4\, \msolar$ is softened (the entropy jump is reduced). With the complete disappearance of the stabilizing gradient provided by Zone~I at $\mathord{\approx}\, 250\, \seconds$, thermal energy can be 
convected into a much larger volume. We define three timescales relevant for the growth of entropy: the convective timescale ($t_{\rm conv}$), the nuclear timescale ($t_{\rm nuc}$), and the expansion timescale ($t_{\exp}$) which are defined as:
\begin{equation}
    t_{\rm conv}\,\mathord{\coloneqq}\,\frac{H_{P}}{\left(\widetilde{v_{\rm r}''^2}\right)^{1/2}},\quad t_{\rm nuc}\,\mathord{\coloneqq}\,\frac{e_{\rm th}}{\dot{q}_{\rm net}},\quad t_{\rm exp}\,\mathord{\coloneqq}\, \left| \frac{\ud \tilde{v}}{\ud r}\right|^{-1},
\end{equation}
where $H_{P}\,\mathord{=}\,P\ud r/\ud P$ is the pressure scale height, and $e_{\rm th}$ is the specific internal energy.
Figure~\ref{fig:figure_11} shows a comparison between the nuclear and convective timescales, $\xi_1\,\mathord{\coloneqq}\,t_{\rm nuc}/t_{\rm conv}$, in the region inside $2.5\,\rm \msolar$ during the last $60\, \seconds$ before the onset of collapse. The value of $\xi_1$ is much larger than unity everywhere (except close to the inner boundary), which means convection can carry away all the nuclear energy released. However, this can not explain the entropy peak seen at the base of the O shell (see the right panel of Figure~\ref{fig:figure_10}). Now, let us take the contraction of the inner core into account. During the last $60\, \seconds$, the inner boundary of the O shell recedes inwards rapidly, while at the same time the outer layers move inward at lower velocities. Therefore, a given mass shell located between $r$ and $r\mathord{+}\delta r$ expands in size. A convective plume which starts at the base of such a shell needs to cover a longer distance to leave the shell. This would result in \textit{trapping} of the convective flow leading to a reduction in the convective efficiency. Figure~\ref{fig:figure_11} shows a comparison of the expansion and convective timescales, $\xi_2\,\mathord{\coloneqq}\,t_{\rm exp}/t_{\rm conv}$. During the last seconds before collapse this ratio approaches unity, which implies that the convective flow is trapped. The inability of convection to transfer the released nuclear energy leads to a rapid buildup of entropy ($\Delta s\,\mathord{\approx}\,0.2\,\kbpnuc$ during the last $5\, \rm \seconds$) at the base of the O shell, leading to an overall negative entropy gradient extending up to $\mathord{\approx}\, 3.6\, \msolar$, which is the base of Zone~II. In the end the 3D model has higher entropy compared to the 1D model.
\subsection{Mixing and Shell Merger}
\begin{figure*}[!]
\begin{center}
\includegraphics{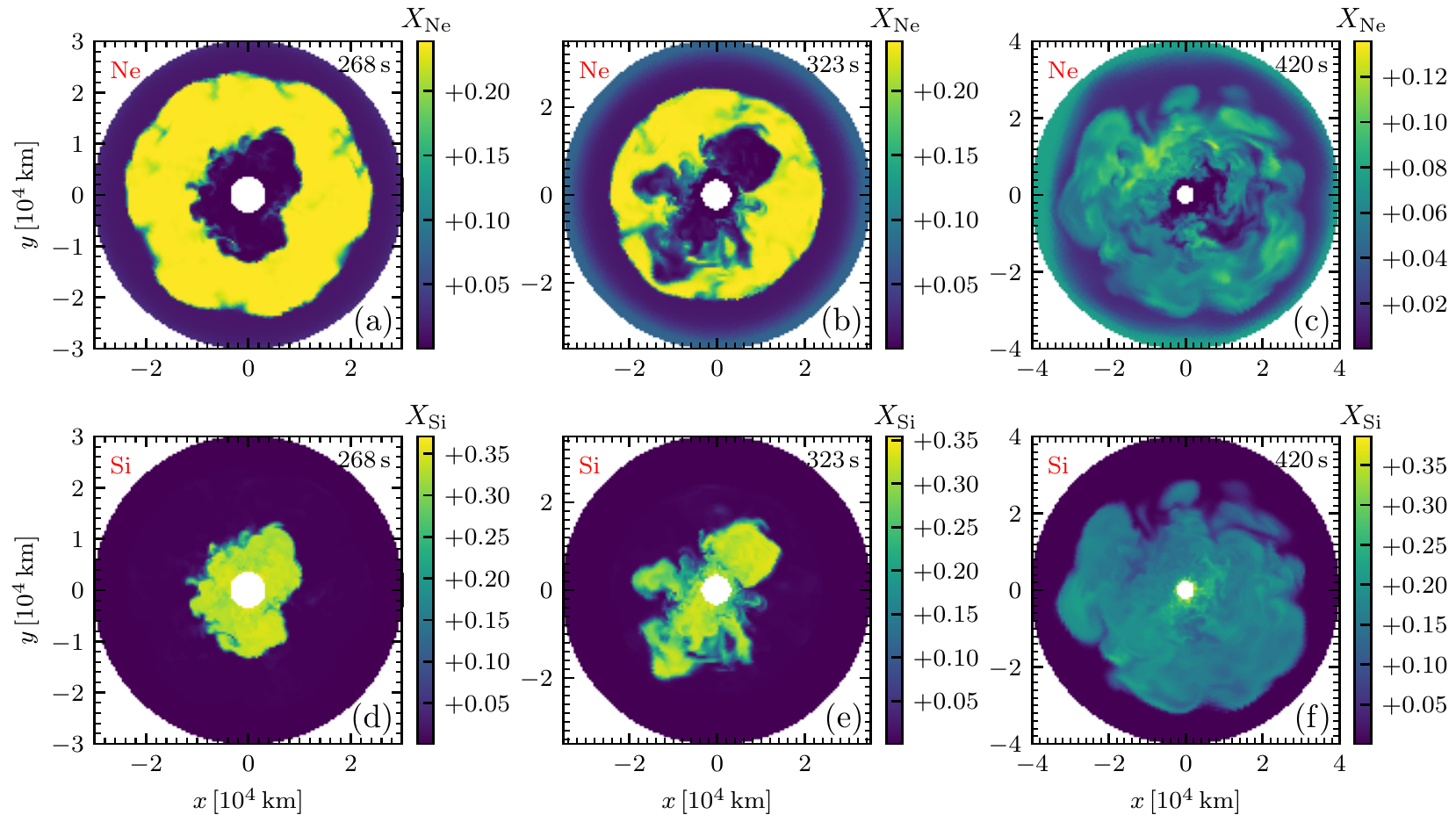}
\end{center}
\vspace*{-6mm}
\caption{Slices ($x\mathord{-}y$ plane) showing the spatial distribution of Ne (upper row) and Si (lower) at various times compared to the volume-rendered 3D views in Figure~\ref{fig:figure_08}. The inner cavity in the Ne distribution (panels~a and b) is shaped by the expanding Si shell (panels~d and e). Panel~b shows prominent Ne downdrafts extending deep into the underlying region and panel~e shows the large-scale Si-rich dipolar plumes. Panels~c and f show the well merged/mixed Ne and Si abundances prior to the onset of gravitational collapse. Each panel shows data only inside a spherical region bounded by a convectively stable layer (Zone~II at $268$, $323\, \seconds$ and Zone~III for $420\, \seconds$) for clarity of presentation. Please see the animation provided as a supplementary material (Movie~B).}\label{fig:figure_12}
\end{figure*}
\begin{figure}[!]
\begin{center}
\includegraphics[scale=0.94]{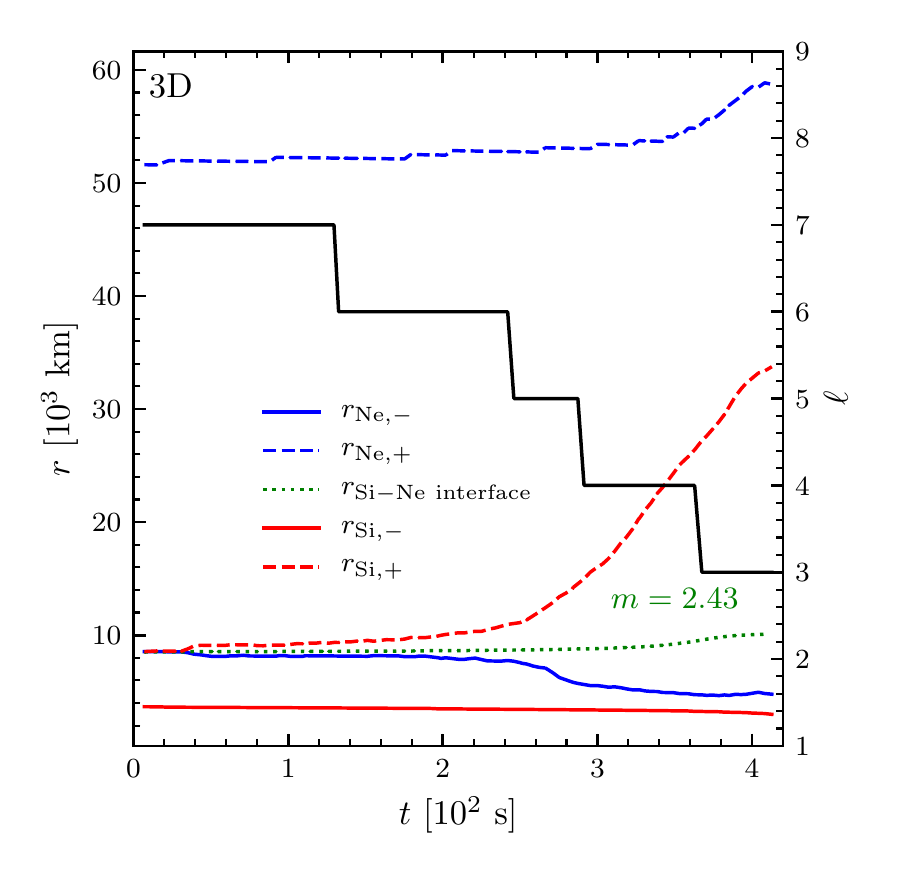}
\end{center}
\vspace*{-7mm}
\caption{Time evolution of the position of the inner boundaries (solid lines, symbol `$-$') and outer boundaries (dashed lines, symbol `$+$') of the Ne and Si layers. The boundaries have been determined using angular averaged mass fraction profiles. The inner boundary of the Ne shell is defined at $X_{\rm Ne}\,\mathord{=}\,0.02$ and the inner boundary of the Si layer is defined at $X_{\rm Si}\,\mathord{=}\,0.40$. The outer boundaries for both Ne and Si are defined at $X_{\rm Ne}\,\mathord{=}\,0.02$ or $X_{\rm Si}\,\mathord{=}\,0.02$, respectively. The evolving radius corresponding to the initial mass coordinate of the Ne/Si interface ($m\,\mathord{\approx}\,2.45$) is shown as the dotted green curve. In the linear regime, the outer boundary of Si shell moves outwards slowly, whereas in the non-linear phase (shell merger) the motion is much faster. The late expansion of the outer Ne shell boundary is caused by an expansion of the whole O-Ne layer due to the violent energy release from enhanced burning. The solid black line shows an estimate of the typical angular mode number of convection cells according to Equation~\ref{eqn:mode} (evaluated for the entire Si-enriched part of the O layer; see last panel of Figure~\ref{fig:figure_12}).}\label{fig:figure_13}
\end{figure}

Figure~\ref{fig:figure_12} shows the spatial distribution of Ne (upper row) and Si (lower row) confirming the picture of large-scale mixing in the 3D model.
The spatial distributions of both Ne and Si at $\mathord{\approx}\, 250\, \seconds$ (panels at $268\, \seconds$) show significant deviations from spherical symmetry. The dipolar feature (panel at $323\, \seconds$) in the Ne distribution is a cavity created by a Si-rich bubble (panel at $323\, \seconds$). The downdrafts of Ne-rich material propagating into the Si-enriched inner volume (panel at $323\, \seconds$) provide the nuclear fuel for \textit{hot-spot} burning. As Ne mixes inwards and burns, the Si-rich matter expands outwards resulting in a complete \textit{merger} of the Ne shell and the Si-enriched oxygen shell as seen in panels at $420\, \seconds$. The Ne mass fraction decreases considerably ($\mathord{\approx}\, 0.2\, \msolar$ of Ne is burnt in total) after the merger and the leftover Ne is thoroughly mixed with Si. Figure~\ref{fig:figure_13} shows an overview of the merger and expansion of the Si and Ne shells. We also show the evolution of the typical angular mode number characterizing convection in the merging layers evaluated using the following relation \citep{Chandrasekhar1961}
\begin{equation}
    \ell \coloneqq \frac{\pi}{2}\times \frac{r_{+}+r_{-}}{r_{+}-r_{-}},\label{eqn:mode}
\end{equation}
where $r_{+}$, $r_{-}$ are the inner and outer radii of the convective shell. The value of $\ell$ decreases rapidly as the ongoing shell merger leads to the formation of a radially extended convective region reaching from the base of the O shell to the outer edge of the convective Ne shell.
\begin{figure*}[!]
\begin{center}
\includegraphics[scale=0.94]{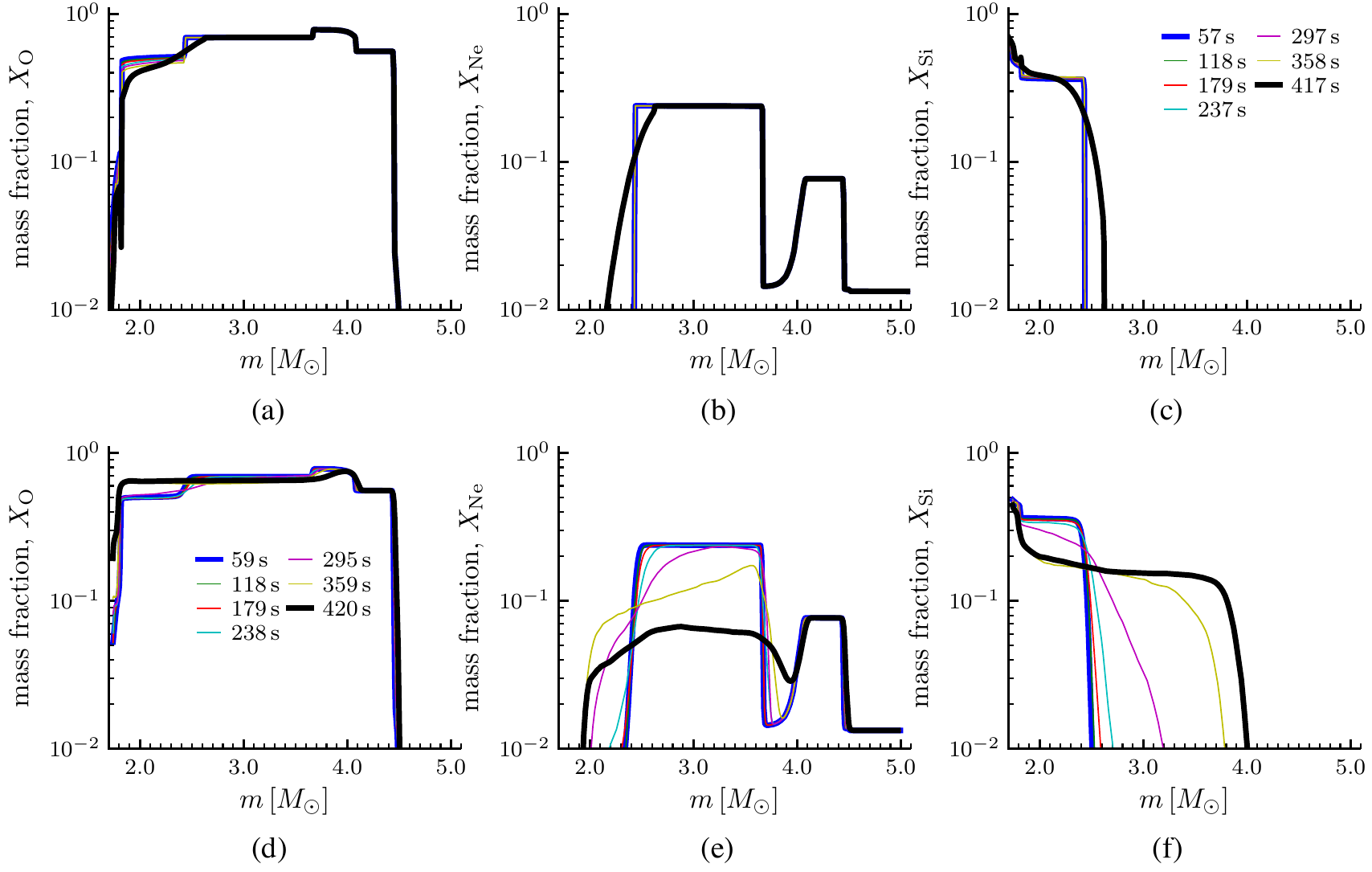}
\end{center}
\vspace*{-7mm}
\caption{Mass fraction profiles for O, Ne, and Si (from left to right) at equally spaced epochs for the 1D (upper row) and 3D (lower row) models. 
The O abundance in the 3D model (panel~d) shows an increase inside $2.5\,\msolar$, whereas such a feature is absent in the 1D model (panel~a). Panel~e shows the sudden decrease in the Ne abundance ($\mathord{\approx}\, 0.2\, \msolar$) in the last $60 \, \seconds$ caused by rapid burning. On the other hand, the Ne abundance in the 1D model (panel~b) is virtually unaffected.
Panel~f shows the outer boundary of the Si-enriched shell reaching out to $\mathord{\approx}\, 35,000\, \kms$ in the 3D model over the course of simulation. In contrast, the 1D model (panel~c) fails to capture the large-scale mixing of Si entirely. Note: The initial abundance profiles are shown in Figure~\ref{fig:figure_01} (top panel).}\label{fig:figure_14}
\end{figure*}
\begin{figure*}[!]
    \begin{center}
    \includegraphics[scale=0.95]{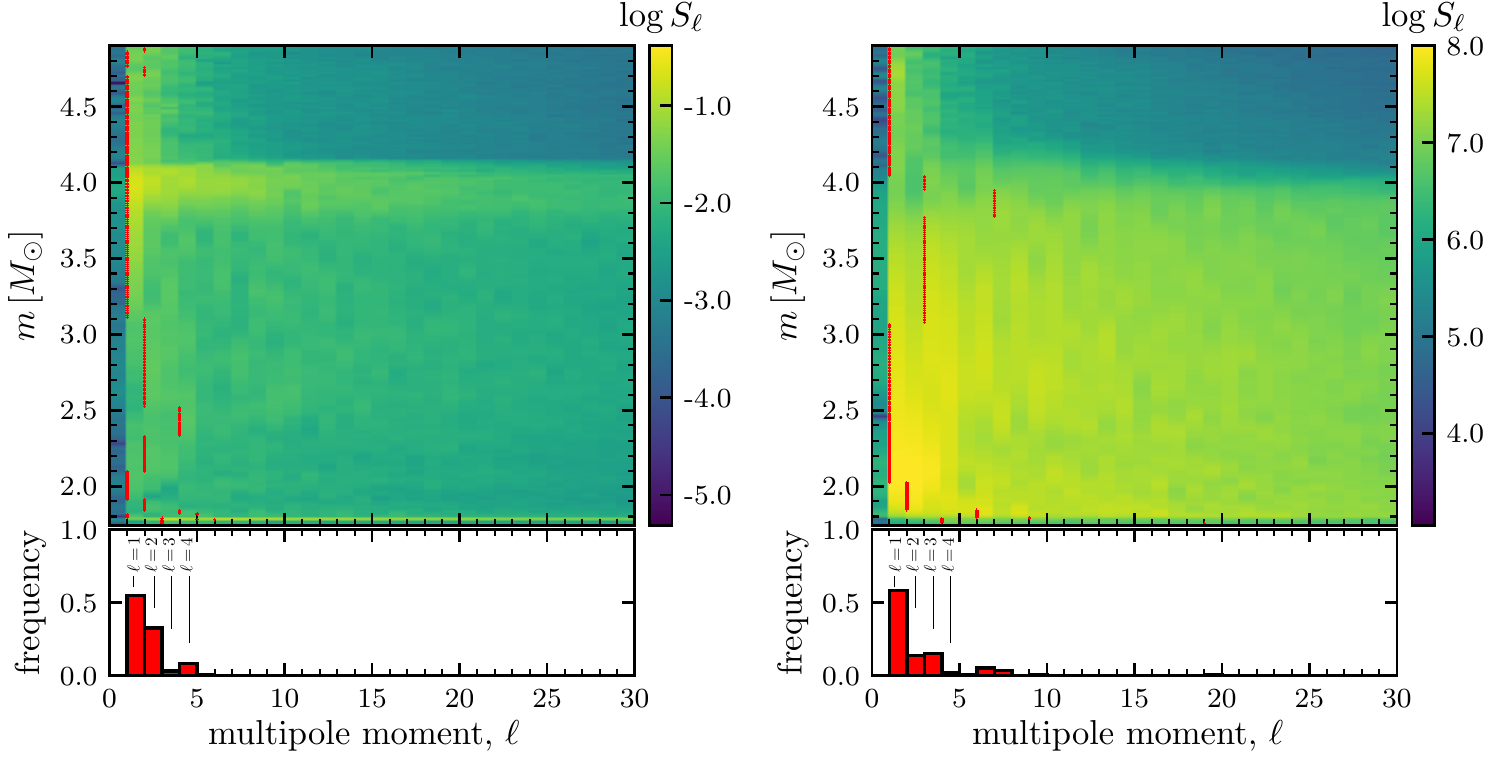}
    \end{center}
    \vspace*{-6mm}
    \caption{Power in various multipoles ($S_{\ell}$) of the density fluctuations ($\rho'/\overline{\rho}$, left panel) and the radial velocity fluctuations ($v''_{r}$, right panel) at the end of simulation. Red marks indicate the mode number with the maximum amplitude at a given mass coordinate. The histogram at the bottom of each panel shows the probability density of multipoles with maximum power in the entire simulated shell. Note: Tick-marks on $\ell$ axis are placed on left edge of a bin, such that the first bin on $\ell$ axis corresponds to $\ell\,\mathord{=}\,0$. Mathematically, the spherical 
    average of a fluctuation (Reynolds averaged) should be zero. In this case, we get a small but non-zero value because of numerical reasons.}
    \label{fig:figure_15}
\end{figure*}

Figure~\ref{fig:figure_14} compares the mass fractions of Si, Ne, and O in the 1D and 3D models. Although both 1D and 3D models show smoothing of the sharp features, the abundances in the 1D model are practically unaffected outside of $\mathord{\approx}\, 2.7\, \msolar$. Abundance changes in the 1D model take place during the last $\mathord{\approx}\, 60\, \seconds$, while in the 3D model they come about slowly over the course of the simulation. In the 1D model these changes results from  convective mixing (MLT), overshooting, thermohaline mixing and semiconvection (treated as diffusive processes, for details refer to \citealt[Section~2.1]{mueller_16c}). Specifically, during the 
last $60\, \seconds$, the core experiences significant contraction ($\approx$50\%). The associated density and temperature increment leads to an enhancement in the neon burning rate, which in turn leads to an increase in the convective velocity resulting in faster mixing (according to the MLT model). In the 3D model they result from mass entrainment at interfaces separating convectively stable and unstable regions, the large-scale convective flow, and the rising Si-rich bubbles powered by rapid Ne burning. 

Panels~a and d compare the O abundance profiles. In the 3D model, the O mass fraction inside $2.5\, \msolar$ increases during the last $60\, \seconds$. The increase results from a combination of large-scale mixing (convective downdrafts carrying O-rich material inwards, $\mathord{\approx}\,0.05 \, \msolar$) and rapid Ne burning (producing O via $^{20}\text{Ne}(\gamma,\alpha)^{16}\text{O}$), and negates the change
caused by burning of O to Si. In the end, the total mass of O in the 
simulated region does not change.

The 1D model fails to capture the evolution of the O abundance altogether. Panels~b and e compare the Ne abundance profiles. The outer part of the neon-carbon shell located between $4.0\mathord{-}4.5\,\msolar$ remains unaffected in both cases. In the 3D model, the gradual rise in the Ne mass fraction inside $2.4\, \msolar$ continues till $\mathord{\approx}\, 360\, \seconds$ only to be reversed by a phase of rapid Ne burning (described in Section~\ref{sec:NucGen3D}). In contrast, Ne is virtually unaffected in the 1D model. Panels~c and f compare the Si abundance profiles. In the 3D model, the evolution of silicon mass fraction is a result of two processes: 1) silicon production by oxygen burning via $^{16}\text{O}(^{16}\text{O},\alpha)^{28}\text{Si}$, and 2) outward mixing caused by convection. During the course of evolution $\mathord{\approx}\, 0.15\, \msolar$ of Si is produced and $\mathord{\approx}\, 0.30\, \msolar$ of Si is gradually transported out from the region between $1.8\mathord{-}2.5\, \msolar$ into the region between $2.5\mathord{-}4.0\, \msolar$. As a result the outer boundary of the Si-enriched layers from $\mathord{\approx}\, 10,000\, \kms$ to $\mathord{\approx}\, 35,000\, \kms$ (close to the C/He interface).
\subsection{Initial Transients and Relaxation}\label{sec:transients}
\begin{figure*}
    \centering
    \includegraphics[width=0.8\linewidth]{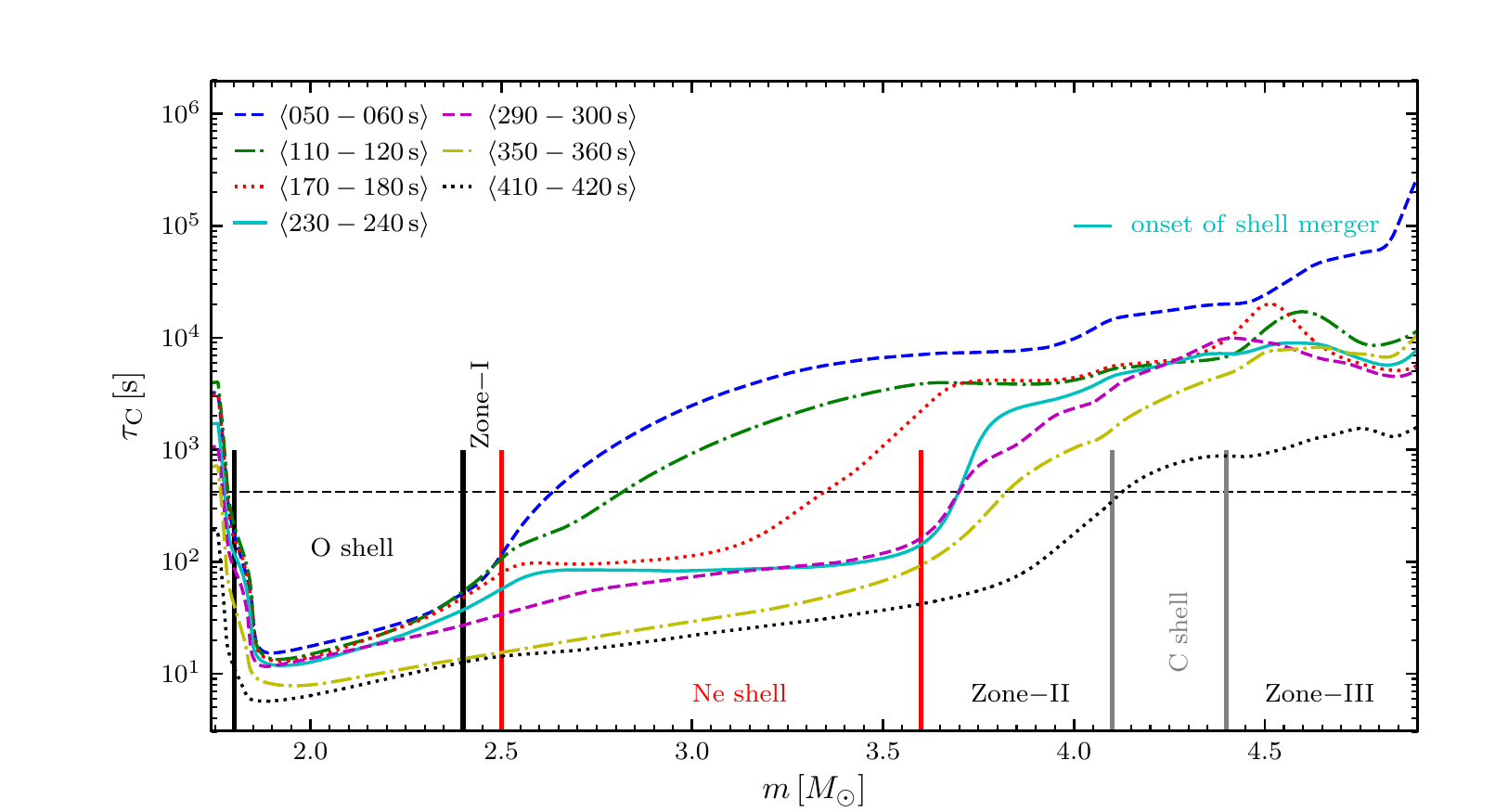}
    \caption{Time averages of convective turnover times (see Equation~\ref{eqn:tauc}) at different instants. The boundaries of the O shell, the Ne shell, and the C shell are marked by black, red, and gray vertical lines, respectively. The curve at $230\, \seconds$ shows that the convective turnover timescale in the layer between the O shell and Zone-I (see Table~\ref{tab:stablezones} for the definition) is $\mathord{\approx}\,10\mathord{-}20\, \seconds$, which gives a conservative estimate of $\mathord{\approx}\,10\mathord{-}20$ convective turnovers in the remaining $190\, \seconds$ before the
    collapse. In the Ne-shell, the convective turnover time is $\mathord{\lesssim}\, 80\, \seconds$, which provides
    enough time for at least $2\mathord{-}3$ convective turnovers in the remaining time before the collapse. The horizontal dashed line marks $420\, \seconds$, which is the total duration of the simulation. The shell merger (described in more detail in Section~\ref{sec:1Dvs3D}) begins at $\mathord{\approx}\, 250\, \seconds$ (shown here by the solid cyan curve).}
    \label{figure_16}
\end{figure*}
Mapping data from one code to another invariably leads to transients. At this point, we must delineate transients caused by mapping errors (\emph{spurious}), and the transients caused by rapid changes in the thermodynamical properties of a system (\emph{genuine}). 

First, we provide two arguments why our results are not affected by mapping errors:
\begin{enumerate}
    \item \textit{Convective Turnover Time}: In order to smooth (damp) mapping related errors, one aims at simulating a sufficiently large number of convective turnovers of the flow. This approach to eliminate the effects of initial transients (spurious) in a simulation is useful when convection is supposed to lead to a steady-state. The convective turnover timescale is conventionally defined as
\begin{equation}
    \tau_{\rm C}\,\mathord{\coloneqq}\,\frac{2H_{P}}{\left(\widetilde{v_{\rm r}''^2}\right)^{1/2}}\label{eqn:tauc},
\end{equation}
where $H_{P}$ is the local pressure scale height, $v_{r}''$ is the fluctuating component of the radial velocity.
Figure~\ref{figure_16} shows the value of the convective turnover timescale averaged over a period of $10\, \seconds$ at various times. The timescale evolves rapidly over the course of the simulation. In the innermost convectively active region, which is the Si-rich part of the O shell (see Figure~\ref{fig:figure_01}), the convective turnover timescale is between $\mathord{\approx}\,10\mathord{-}20\, \seconds$ during the entire simulation, which is sufficient for $20\mathord{-}40$ full turnovers during the simulation. In the Ne shell, the turnover timescale becomes relatively stable at around $230\, \seconds$ and its value is $\mathord{\lesssim}\, 80\, \seconds$ thereafter. Therefore, even after the onset of the shell merger, which begins at $\mathord{\approx}\, 250\, \seconds$ (see section~\ref{sec:1Dvs3D}), the region has enough time to go through 
$2\mathord{-}3$ convective turnovers. Furthermore, the convectively stable layers sandwiched between the convectively unstable layers (see Section~\ref{subsec:conv-stab}) play an effective role in
absorbing and dissipating the initial transients because they damp the propagation of acoustic waves.

\item \textit{Growth of Kinetic Energy In the Merging Shells}:
During the onset phase, between $0\,\mathord{-}\, 80\, \seconds$ (Figure~\ref{fig:figure_09}), the kinetic energy in radial velocity fluctuations grows and reaches a steady value in the layers of interest, which are the Si-rich part of the O shell (labelled as ``convective O shell'') and the convective Ne shell, implying that most of the initial transients (mapping errors) have already dissipated away and do not affect the further evolution. 
\end{enumerate}

In the case of a genuine transient, the system may undergo an irreversible change in a short time, and may never approach a steady-state, which is also the case with the shell-merger described above. In this case, the numerical solution is stable (the CFL value is 0.6; the code preserves the hydrostatic state; PPM is well understood), but the physical system itself is on the verge of a thermodynamical instability. The fact that the system under consideration here manifests such a behaviour is evident from the nature of the transition between the `stationary' phase and the `rapid rise' phase (see Figure~\ref{fig:figure_09}). Before this transition, the kinetic energy in the convective O shell and Zone-I stays at a plateau between $\mathord{\approx}100\,\mathord{-}\, 300\, \seconds$, and in the Ne-shell energy is slowly deposited by convective activity underneath and gently reaches a steady state value at $\mathord{\approx}\, 200\, \seconds$. For the next
$\mathord{\approx}\, 125\, \seconds$ the change in kinetic energy in the Ne shell is negligible. 
Its value starts rising again only at the time of the transition in response to the elevated energy production in the O burning shell, associated with the shell merger beginning at $\mathord{\approx}\, 250\, \seconds$. Therefore, the ensuing shell merger represents a ``genuinely transient phenomenon'' consistent with the thermodynamical evolution. It is obvious that the conditions emerging from the shell merger do not represent a steady state situation, and we cannot claim that the same conditions or the same dynamical situation would be present at the onset of collapse if we had started our simulation significantly earlier. However, we consider the shell merger starting $\mathord{\approx}\, 200\, \seconds$ before collapse as an exemplary case of such a possibility, which could happen in other stars even if not in the one considered here. In Section~\ref{sec:Summary}, we further discuss the effects of our choice of the starting time on the onset and dynamics of the shell merger and its impact on the collapse conditions.
\section{Pre-Supernova Model Properties}\label{sec:pre-sup}
\begin{figure*}[!]
    \begin{center}
    \includegraphics[scale=1.0]{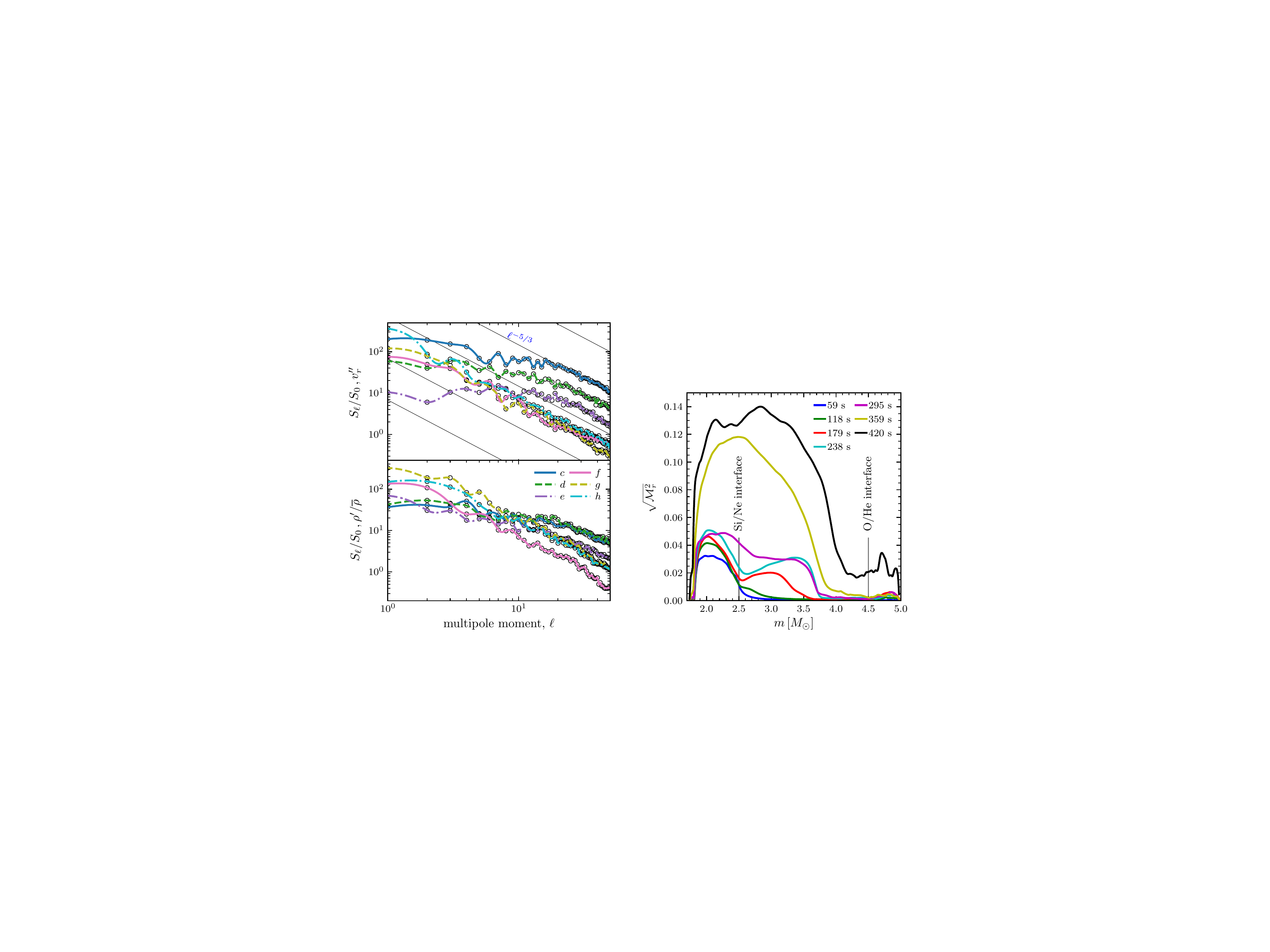}
    \end{center}
    \vspace*{-6mm}
    \caption{Left panels: Multipole power spectra ($S(\ell)/S(0)$) for density fluctuations (lower half-panel) and radial velocity fluctuations (upper half-panel) at the center of the various layers (\textit{c, d, e, f, g}, and \textit{h}) defined in Section~\ref{sec:corr_len} at the end of simulation. Black circles mark the computed harmonics (integer values only) and the smooth curves are cubic splines. The black lines in the upper half-panel indicate a slope of $-5/3$ expected for the inertial range in a fully developed turbulent flow according to Kolmogorov's theory. Right panel: Root mean square Mach number of the radial velocity fluctuations. The Si/Ne interface is at the boundary of layer c and layer d, and the O/He interface is at boundary of layer g and layer h. Up to about $300 \, \seconds$, distinct convective O and Ne shells can be recognized, both with strongly subsonic convection. After the shell merger, the convective Mach numbers increase considerably.}
    \label{fig:figure_17}
\end{figure*}

Large-scale modes with $\ell \mathord{\sim} 1\mathord{-}2$ are most effective for shock revival \citep{mueller_15a,abdikamalov_16,mueller_16a}. In order to gain a quantitative understanding of the modes, we have done spherical harmonic decomposition of the density and velocity perturbations. Figure~\ref{fig:figure_15} shows the multipole power distribution at collapse for the density fluctuations (left panel) and the velocity fluctuations (right panel). The red markers indicate the multipole with the largest power ($\lmax$) at a given mass coordinate. The value of $\lmax$ changes with radius, but there are continuous sections where it remains stationary. In the case of density fluctuations inside the merged O-Ne shell ($1.8\mathord{-}3.1\, \msolar$) a mixture of dipole ($\ell\,\mathord{=}\,1$) and quadrupole mode ($\ell\,\mathord{=}\,2$) dominates, and in the region outside of Zone~II the dipole mode dominates (similar to the global non-spherical oscillation described by \citet{herwig_14}). This confirms our visual impression obtained from Figure~\ref{fig:figure_03}. In the case of radial velocity fluctuations a mixture of dipole mode and $\ell=3$  mode dominates in the merged O-Ne shell, whereas the dipole mode dominates in the C and He shells. The normalized probability distribution of $\lmax$ (shown as histograms) confirms that $\ell\,\mathord{\sim}\,1\mathord{-}3$ dominate both the density fluctuations and the velocity fluctuations. Figure~\ref{fig:figure_17}, left panels, show another view of the spectra at the middle points of the layers defined in Section~\ref{sec:corr_len}. In the lower half-panel the spectra for the region inside $\mathord{\approx}\, 3.6\, \msolar$ (layers~c and d) are shallow compared to the region outside and including Zone~II. In the upper half-panel the spectra for $\ell\,\mathord{\lesssim}\,15$ are shallower in the merged O-Ne shell including Zone~II. The spectra above $\ell\,\mathord{\gtrsim}\,15$ agree quite well with the `Kolmogorov slope'.

Figure~\ref{fig:figure_17} (right panel) shows the profiles of the turbulent Mach number corresponding to the radial velocity fluctuations at multiple times. The quantity is defined following \citet{mueller_16c} as
\begin{equation}
    \widetilde{\mathcal{M}^2_{r}} \coloneqq \dfrac{\int\limits_{\Omega} \rho(v_{r}-\tilde{ v}_{r})^2\, \ud\Omega}{\int\limits_{\Omega} \rho c_{\text{s}}^2\, \ud\Omega},
\end{equation}
where $c_{\text{s}}$ is the adiabatic sound speed. The Mach number in the O and Ne shells increases gradually until the shell merger. During the first $120\, \seconds$, the value of $\mathcal{M}_{r}$ is almost negligible in the Ne shell outside $2.5\, \msolar$, which is due to the smaller Brunt-V{\"a}is{\"a}l{\"a} in the Ne shell. The region outside $4.1\, \msolar$ stays quiescent until $360\, \seconds$ as the stable Zone~II located between $3.6\mathord{-}4.1\, \msolar$ does not allow convective plumes to penetrate further. After the merger, between $300\mathord{-}360\, \seconds$, $\mathcal{M}_{r}$
increases rapidly by a factor of $\mathord{3}$ reaching a peak value of $\mathord{\approx}\, 0.14$ at the onset of collapse. Convection is so violent that the plumes penetrating deep into Zone~II excite strong interfacial waves that reach as far as the He shell, visible by the non-negligible Mach numbers outside $4.5\, \msolar$. 
\begin{figure*}[!]
    \begin{center}
        \includegraphics[width=1.0\linewidth]{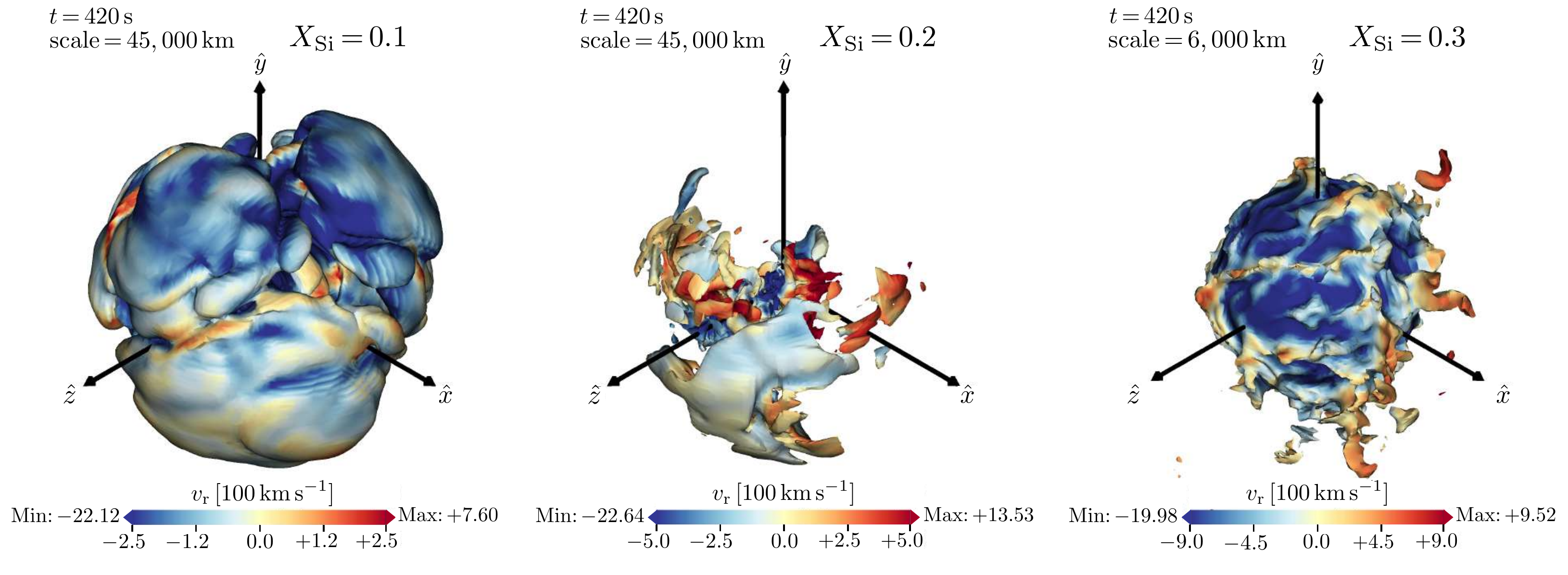}
    \end{center}
    \vspace*{-6mm}
    \caption{Pseudocolor maps of radial velocity $v_{\text{r}}$ on Si mass fraction isosurfaces for $X_{\text{Si}}\,\mathord{=}\, 0.1$ (left), $X_{\text{Si}}\,\mathord{=}\,0.2$ (middle) and $X_{\text{Si}}\,\mathord{=}\,0.3$ (right), which effectively probe the Si distribution at successively greater depths. The snapshots are taken at $420\, \seconds$ when the Fe-core begins to undergo gravitational collapse. ``Scale'' refers to the size of the $y$-axis (vertically upwards). The left panel shows the large-scale Si-rich bubbles. The middle panel shows the asymmetry and clumpy Si distribution prior to collapse, and the right panel shows Si-rich plumes and downdrafts in the deep interior moving outward or inward, respectively, at $\mathord{\approx}\, 1,000\, \kmps$. Please see the animation provided as a supplementary material for the case with $X_{\text{Si}}\,\mathord{=}\, 0.1$ (Movie~C).}
    \label{fig:figure_18}
\end{figure*}
Figure~\ref{fig:figure_18} shows pseudocolor maps of the radial velocity on different Si mass-fraction isosurfaces prior to collapse. As we go from small to large values of $X_{\text{Si}}$, we effectively probe the Si distribution in successively deeper regions of the star. The left panel shows large-scale Si-rich bubbles rising at moderate speeds of $\mathord{\approx}\, 100\, \kmps$ somewhere around $\mathord{\approx}\, 30,000\mathord{-}40,000\, \kms$. In the panel for $X_{\text{Si}}\, \mathord{=}\, 0.2$ (middle), we see a highly asymmetrical and clumpy distribution of Si with some of the clumps moving outwards at velocities in excess of $1,000\, \kmps$. This may have an impact on the observed asymmetries in the element distributions in some young supernova remnants such as Cassiopeia~A. In the panel for $X_{\text{Si}}\, \mathord{=}\, 0.3$ (right) we probe an even higher Si mass fraction which is present deep inside the O shell (scale of $\mathord{\approx}\, 6,000\, \kms$), where we again see fast outward moving Si-rich plumes. In short, the Si distribution in bulk is clumpy and has asymmetric  velocities (although it looks as if it had isotropized in panel~f of Figure~\ref{fig:figure_12}).
\section{Conclusions}\label{sec:Summary}
In this work, we present the first 3D simulation (full $4\pi$ solid angle) of a fully developed large-scale O-Ne shell merger prior to core-collapse in an $18.88\, \msolar$ progenitor. The work builds upon the previous study of \citet{mueller_16c} in which they simulated the last minutes of O-shell burning in an $18\, \msolar$ supernova progenitor up to the onset of core collapse. In the present case we find significant differences between the kinematic, thermodynamic, and chemical evolution of the 3D and the 1D models. Ne is mixed deep into the O-shell, leading to a fuel ingestion episode which 
releases significant amounts of thermal
energy by nuclear burning
on a short timescale ($\mathord{\sim}100\, \seconds$), further boosting convection:
Convective downdrafts transport Ne close
to the base of the O shell, where it ignites and powers convection in return, leading to a positive feedback loop for rapid Ne entrainment.
The entire simulated shell attains a convective Mach number of $\mathord{\approx}\, 0.1$. The maximum convective velocities are $\mathord{\sim}\, 1300 \, \kmps$ for updrafts and $\mathord{\sim}\, 1700 \, \kmps$ for downdrafts. In contrast the 1D model is relatively quiescent with convective shell burning constrained to layers. The specific kinetic energy in the merged shell increases by a factor of $\mathord{\approx}\, 50\mathord{-}100$ during the course of evolution. Although the Si-rich material forms a strongly dipolar structure in the merged shell at $(\mathord{\approx}\, 330\, \seconds)$, the distribution of elements becomes more isotropic in the following time leading up to collapse. However, the velocity field still shows a large global asymmetry at collapse. In short, the 3D pre-supernova model exhibits strong density perturbations and large-scale velocity asymmetries (with dominant $\ell\mathord{\sim} 1\mathord{-}2$ modes) in the flow as well as the distribution of nuclear species.

Another question worth investigating is that, ``how the \textit{choice} of starting time affects the onset and dynamics of the shell merger?'' Although, we can definitely say that the shell merger reported in this study is 
a robust phenomenon, the exact time at which it happens may perhaps change if we start at a different time. 
This question can only be answered by doing a longer simulation, which is at the moment 
unfeasible due to limited computing resources. Nevertheless, the shell merger \emph{must} occur quite
late during the evolution of the O shell because this is when
the critical conditions (a reduction of the entropy jump 
between the O shell and Ne shell, and fast O shell convection
that allows fast entrainment) are established. This point has been
discussed in detail in \citet{collins_18}: For the entropy
of the O shell to rise and become similar to that of the
Ne shell, O burning and neutrino cooling must have already
dropped out of balance due to the contraction of the
O shell. The contraction of the O shell in the wake
of core contraction allows convection to reach sufficiently high Mach numbers
and power strong entrainment.

Simulations by \citet{couch_15} and \citet{mueller_16c} have already shown that the progenitor structure is genuinely three-dimensional at collapse, but only considered quasi-steady state convection. 
Recently, \citet{Yoshida_2019} have presented a suite of 2D and 
3D models of O-shell burning for massive stars between $9\mathord{-}40\, \msolar$. They find convection with large-scale eddies and the turbulent Mach number $\mathord{\sim}\, 0.1$ in their models despite the simulations being short (lasting $\mathord{\approx}\, 100\,\seconds$ before collapse), which nevertheless shows that large-scale perturbations may be 
a generic feature in pre-supernova stars. Our work demonstrates that asymmetries can become exceptionally large when the convective flow becomes non-steady because of a shell merger. In such a case, the nuclear timescale becomes comparable to the convective timescale, and hence nuclear burning becomes strongly coupled with the flow dynamics and the resultant mixing is considerably faster and can be highly asymmetric. 1D stellar evolution calculations are severely limited in capturing these phenomena self-consistently as they occur primarily due to a combination of turbulent entrainment, fuel ingestion, and the excitation/propagation of internal waves, which are inherently 3D effects. Despite the inherent limitations of 1D calculations, such shell mergers may still take place in 1D (as seen by \citet[model S20]{rauscher_02} and in recent studies by \citet{sukhbold_14} and \citet[case C3]{davis_2018}, and merely be delayed as in model S25 of \citet{rauscher_02}, where the shell merger happens only $5\, \seconds$ before collapse. However, shell mergers in 1D models cannot capture the highly dynamical and asymmetric flow during and after the merger. Therefore, the dynamics of shell mergers during the late burning stages can only be captured using self-consistent multi-dimensional calculations. Such calculations are still in their infancy and the work reported here attempts to make progress in our understanding of the internal structure of supernova progenitors. 

Also, recent studies show that progenitor asymmetries may help in shock revival \citep{couch_13,mueller_15a,mueller_17} by aiding the growth of convection and/or instabilities like the standing accretion shock instability (SASI) \citep{takahashi_16,mueller_17}. Our work furnishes further evidence that some supernova progenitors have convective seed perturbation that are dynamically important. For a shell merger like the one described in this work, we see even more violent convection and stronger large-scale asymmetries than in the $18.0\, \msolar$ model of \citet{mueller_17}. 

Shell mergers will also affect the compactness parameter \citep{oconnor_11} and thus the explodability with possible consequences for the location and extent of ``islands of explodability'' (e.g. \citet{sukhbold_14}). Due to the high convective Mach numbers and pronounced
global asymmetries, we expect a strong impact on the explosion dynamics for our $18.88\, \msolar$ model. 

This work constitutes only the first step towards understanding the importance of late-stage shell mergers. For example, mergers may also impact
the pre-supernova nucleosynthesis yields, as already suggested by 1D stellar evolution models.
The S20 model of \citet{rauscher_02} showed a strong overproduction of several elements between Si and V and an underproduction of Cr, Mn, and the light Fe isotopes, which they ascribed  to the stellar structure resulting from the C/O shell merger.
\citet{ritter_18} showed that even moderate C-ingestion events during O burning can significantly overproduce odd-Z elements such as K, Sc, Cl, etc. Some of their models with full C-O shell merger (albeit artificially increased O burning luminosity) have overproduction factors greater than 10. Therefore, we expect to see substantial effects on the nucleosynthetic yields due to the atypical way in which Ne burns during the shell merger event described in our work. During the explosion phase, the substantial asymmetries in the element distribution and the velocity field  may have  repercussions for observable asymmetries in supernovae and nucleosynthesis \citep{hughes_2000,delaney_10,lopez_2018}. Such questions will be addressed in the future.

\section*{Acknowledgements}
This project was supported by the European Research Council through grant
ERC-AdG No.\ 341157-COCO2CASA and by the Deutsche Forschungsgemeinschaft (DFG, German Research
Foundation) under Germany's Excellence Strategy through 
Excellence Cluster ORIGINS (EXC-2094)--390783311.
This work was also supported by the Australian Research Council through
ARC Future Fellowships FT160100035 (BM), Future Fellowship
FT120100363 (AH). This research was undertaken with the assistance of
resources obtained via NCMAS and ASTAC  from the National Computational Infrastructure (NCI), which
is supported by the Australian Government and was supported by
resources provided by the \textsc{Pawsey Supercomputing Centre} with funding
from the Australian Government and the Government of Western
Australia. On the Garching side, computing resources from the \textsc{Gauss Centre for Supercomputing} (at the \textsc{Leibniz Supercomputing Centre} (LRZ) under SuperMUC project ID: pr53yi) are acknowledged. The analysis of the simulation was done using the MPG Supercomputer \textsc{Hydra}, \textsc{Cobra}, and \textsc{Draco}
provided by \textsc{The Max Planck Computing and Data Facility} at Garching. 
\software{NumPy \citep{Walt_2011}, Scipy \citep{oliphat_2007},
IPython \citep{PER-GRA:2007},
Matplotlib \citep{Hunter:2007}, Visit \citep{Childs_11}, and SHTns\, library \citep{schaeffer_2013}.}
This research has made use of NASA's Astrophysics Data System.
\appendix

\section{Angle Averaging Method}\label{sec:favre-avg}
\noindent
An intensive physical quantity (for e.g., density) can be written as a sum of its average and the fluctuating component (notation for mean and fluctuating components follows from \citet[chapter~5]{Oertel_2010}) as $f(\mathbf{r}) = \overline{f} + f'(\mathbf{r})$, where the angle-averaged value (Reynolds) $\overline{f}$ is defined as
\begin{equation}
\overline{f}\coloneqq \langle f \rangle = \frac{1}{4\pi}\int\limits_{\Omega} f(\mathbf{r})\, \ud \Omega.
\end{equation}
An extensive physical quantity (for e.g., specific energy $e$) can be written as $
g(\mathbf{r}) = \tilde{g} + g''(\mathbf{r})$, where the angle-averaged value (Favre) is defined as
\begin{equation}
\tilde{g} \coloneqq \langle g \rangle = \dfrac{\int\limits_{\Omega} \rho(\mathbf{r})g(\mathbf{r})\,  \ud\Omega}{\int\limits_{\Omega}\rho(\mathbf{r})\, \ud \Omega}.
\end{equation}
The fluctuating components $f'$ and $g''$ satisfy the following relations
\begin{equation}
\overline{f'} = \frac{1}{4\pi}\int\limits_{\Omega} f'(\mathbf{r})\, \ud\Omega = 0,  \quad\quad
\overline{g''}= \frac{1}{4\pi}\int\limits_{\Omega} g''(\mathbf{r})\, \ud\Omega \ne 0,\quad \text{and},\quad\quad 
\overline{\rho g''} = \frac{1}{4\pi} \int\limits_{\Omega} \rho(\mathbf{r})g''(\mathbf{r})\, \ud\Omega = 0.
\end{equation}
\section{Spherical Harmonic Transforms}\label{sec:shts}
\noindent
The forward spherical harmonic transform of a function 
$f(\theta,\phi)$ involves computing the coefficients
$c_{\ell}^m$ according to
\begin{equation}
    c_{\ell}^{m} = \int\limits_{\Omega} {Y_{\ell}^m}^{\ast}(\theta,\phi) f(\theta,\phi) \, \ud \Omega,
\end{equation}
where $Y_{\ell}^m(\theta, \phi)$ is the orthonormalized spherical harmonic function of degree $\ell$ and order $m$. The total power $c_{\ell}^2$ for a mode with multipole order $l$ is defined as
\begin{equation}
S_{\ell} \coloneqq c_{\ell}^2 = \sum_{m=-\ell}^\ell \left |\int\limits_{\Omega} Y_{\ell}^{m} (\theta,\phi)
f(\theta,\phi) \, \ud \Omega\right|^2.
\end{equation}
We have used the routines from the SHTns library \citep{schaeffer_2013} to calculate the spherical harmonic decomposition.
\bibliography{paper}
\bibliographystyle{aasjournal}
\end{document}